\documentclass[nonacm=true, sigconf]{acmart}

\usepackage{epsfig,hyperref,color,soul,endnotes,listings,soul,subfig,multirow,tabularx,graphics,epstopdf}

\begin{document}
\title{Catch Me (On Time) If You Can:\\Understanding the Effectiveness of Twitter URL Blacklists}

\author{Simon Bell}
\affiliation{
  \institution{Royal Holloway, University of London}
}
\email{simon.bell.2014@rhul.ac.uk}

\author{Kenny Paterson}
\affiliation{
  \institution{ETH Zurich}
}
\email{kenny.paterson@inf.ethz.ch}

\author{Lorenzo Cavallaro}
\affiliation{
  \institution{King's College London}
}
\email{lorenzo.cavallaro@kcl.ac.uk }

\date{\today}

\fancyhf{} 
\fancyhead[C]{Catch Me (On Time) If You Can: Understanding the Effectiveness of Twitter URL Blacklists}
\fancyfoot[C]{\thepage}

\setcopyright{none} 
\settopmatter{printacmref=false} 

\renewcommand\footnotetextcopyrightpermission[1]{} 

\begin{abstract}
With more than 500 million daily tweets from over 330 million active users, Twitter constantly attracts malicious users aiming to carry out phishing and malware-related attacks against its user base. It therefore becomes of paramount importance to assess the effectiveness of Twitter's use of blacklists in protecting its users from such threats. We collected more than 182 million public tweets containing URLs from Twitter's Stream API over a 2-month period and compared these URLs against 3 popular phishing, social engineering, and malware blacklists, including Google Safe Browsing (GSB). We focus on the delay period between an attack URL first being tweeted to appearing on a blacklist, as this is the timeframe in which blacklists do not warn users, leaving them vulnerable. Experiments show that, whilst GSB is effective at blocking a number of social engineering and malicious URLs within 6 hours of being tweeted, a significant number of URLs go undetected for at least 20 days. For instance, during one month, we discovered 4,930 tweets containing URLs leading to social engineering websites that had been tweeted to over 131 million Twitter users. We also discovered 1,126 tweets containing 376 blacklisted Bitly URLs that had a combined total of 991,012 clicks, posing serious security and privacy threats. In addition, an equally large number of URLs contained within public tweets remain in GSB for at least 150 days, raising questions about potential false positives in the blacklist. We also provide evidence to suggest that Twitter may no longer be using GSB to protect its users.
\end{abstract}

\ccsdesc[500]{Security and privacy~Malware and its mitigation}
\ccsdesc[500]{Security and privacy~Phishing}
\ccsdesc[300]{General and reference~Empirical studies}
\ccsdesc[300]{General and reference~Measurement}

\keywords{Security, Malware, Phishing, Blacklists, Measurement Study}

\maketitle

\section{Introduction}

Since its creation in 2006, Twitter has gained over 974 million users
with 330 million active users per month posting 500 million tweets per day~\cite{twitter-stats}. Among these Twitter users are many high profile celebrities, politicians, heads of state and societal influencers whom attract large numbers of followers~\cite{twitter-counter}. Due to this large user base, Twitter makes an attractive target for malicious users aiming to carry out phishing and malware attacks to exploit people. One of the main ways these attacks are carried out is by leading victims to a malicious site, by including one or more URLs in a tweet, whereby the attack can occur.

Phishing attacks on Twitter have been known to lure victims in by
offering verification on the social network but instead take them to a
fake login page to steal their Twitter username and
password~\cite{daily-dot}, while malware attacks have included
drive-by-download links contained within tweets, cross-site scripting
attacks~\cite{securelist}, and Android malware that is controlled by tweets~\cite{eset}.

Twitter has come under increasing pressure to protect its users against these attacks, such as, in 2010 when the company settled a case with the US Federal Trade Commission in which Twitter agreed to strengthen security throughout the platform and to carry out an independently assessed bi-annual information security audit~\cite{FTC}.

One of the ways in which Twitter is improving its security for users is by implementing numerous rules that govern what type of content users of the platform can and cannot send~\cite{twitter-rules}. In 2009, it was reported~\cite{zdnet} that Twitter had started to use the phishing and malware blacklist Google Safe Browsing (GSB), already used by popular web browsers to filter out and protect its users from attack URLs. We provide evidence that suggests Twitter is not using GSB effectively to protect its users.

Our paper aims to assess how effective Twitter's use of blacklists is in protecting its users from phishing and malware attacks. In particular, we focus on the delay period between an attack URL first being tweeted to appearing in one of 3 defined blacklists, as this is the timeframe in which blacklists do not warn users against the attack. 
We collected over 182 million public tweets containing URLs from Twitter's Stream API over a 2-month period and compared these URLs against 3 popular phishing, social engineering, and malware blacklists that are used in leading web browsers, antivirus solutions, and other online protection technologies. 

During one month we discovered 4,930 tweets containing URLs leading to social engineering websites that had been tweeted to over 131 million Twitter users. The majority of URLs contained within these tweets took between 20 and 30 days to appear in GSB. We focus on GSB because it is the main protection used in popular web browsers. In the same month we also discovered 1,126 tweets containing 376 blacklisted Bitly URLs that had a combined total of 991,012 clicks -- these Bitly URLs represent 11\% of the total blacklisted social engineering URLs in our dataset for that month. This demonstrates that Twitter users are clicking on and being exposed to dangerous websites. 

We also discovered that, while the GSB blacklist is effective at blocking a large number of social engineering and malicious URLs within 6 hours of being tweeted, a large number of URLs go undetected for at least 20 days, with users potentially exposed to attacks during this delay. In addition, an equally large number of URLs contained within public tweets remained in the GSB blacklist for at least 150 days, potentially raising issues with false positives in the blacklist.

Twitter provides a Stream API to access a source of live tweets. There are 3 ways of accessing this API: the filter/sample, decahose, and firehose streams. These feeds contain, approximately, 1\%, 10\%, and 100\% of all public tweets, respectively. The filter/sample feed is free to access, while the decahose and firehose feeds come at a substantial cost. Our study made use of Twitter's filter/sample stream. There are methodological limitations to using this smaller sample feed. For example, URLs of interest may not be contained in the feed we receive. We compensate as much as possible for this, with techniques such as using Twitter's Search API to determine original tweet date instead of relying entirely on what our 1\% sample tells us.

To the best of our knowledge ours is the first in-depth study that specifically focuses on the impact of blacklist delays on Twitter traffic. Our study provides a present-day snapshot of the current state of phishing and malware URLs being posted to Twitter. A previous study from 2010~\cite{grier2010spam} took important first steps in this direction, but Twitter's active user base has grown from 30 million users in 2010 to 330 million users in 2017~\cite{statistica} and the number of daily tweets has grown from 35 million in 2010 to over 500 million in 2017~\cite{InternetLiveStats}. We replicate the  experiment of~\cite{grier2010spam} (to the extent we can in the face of missing details in~\cite{grier2010spam}) but also present a more comprehensive and detailed analysis of malicious URLs on Twitter. In particular, we introduce a new methodology to measure delay from first tweet to membership in the GSB blacklist to determine effectiveness of Twitter URL blacklists. We are also able to determine worst-case scenario delay periods, and we measure the duration of time that URLs stay in GSB.

We organise the remainder of this paper into the following sections:
Section~\ref{sec:background} introduces the background and related
work, Section~\ref{sec:design} describes the design and infrastructure
we used to carry out experiments and the experiments themselves,
Section~\ref{sec:implementation} describes how we implemented the
infrastructure and experiments, Section~\ref{sec:results} presents our
results, Section~\ref{sec:discussion} discusses the main findings, and
Section~\ref{sec:conclusion} provides concluding remarks.

\section{Background and Related Work}
\label{sec:background}

\subsection{Blacklists}

A blacklist is defined as a set of elements to be blocked; an access control list. Our study looks at phishing and malware blacklists that are used to block access to URLs posted to Twitter. We focus on 3 blacklists: Google Safe Browsing~\cite{safebrowsing}, Open Phish~\cite{openphish}, and Phish Tank~\cite{phishtank}.

\paragraph{Google Safe Browsing:} Google Safe Browsing (GSB) is a URL blacklist that contains both malicious and phishing URLs and is used by the web browsers Google Chrome, Safari, Firefox, Opera, and Vivaldi to protect users from dangerous websites. We focus on GSB in our study because of its prominence in popular web browsers: already in 2012 GSB was protecting 600 million users from dangerous websites~\cite{webpronews}. In 2015 GSB began using the term ``Social Engineering'' to categorise phishing websites which also encompass additional types of deceptive content. Google defines a social engineering web attack as occurring when either: ``the content pretends to act, or looks and feels, like a trusted entity - like a bank or government'' or ``the content tries to trick you into doing something you would only do for a trusted entity - like sharing a password or calling tech support''~\cite{google-security-blog}. During the week commencing 3rd September 2017 the total number of sites deemed dangerous by GSB was 573,433 phishing and 500,245 malicious. During that week GSB detected 24,756 new phishing sites and 6,312 new malware sites. GSB defines malware websites in its blacklist as being either \emph{compromised} or \emph{attack}. A compromised website is a legitimate website that has been hijacked to either include, or direct users to, malicious content. An attack site is a website that has intentionally been set up to host and distribute malware~\cite{google-transp-report}. During the week commencing 3rd September 2017, GSB identified 5,981 new compromised websites and 335 new attack websites.

GSB provides two APIs for accessing its blacklist: Lookup and Update. The Lookup API provides a remote service whereby URLs to be checked are sent to Google's servers and a response is returned for each URL stating if the URL is in the blacklist. The Update API provides the user with a local copy of the blacklist, this local copy is stored as a database of SHA-256 URL hash prefixes, the majority of the hash prefixes being 4 bytes. To perform a URL blacklist lookup, the URL hash prefix is checked in the local database and, if there is a prefix match, then the full URL hash is retrieved from Google's servers to determine if there is a match on the full hash.

\paragraph{Open Phish:} Open Phish launched in 2014 and is the result of a 3 year research project on phishing detection that uses autonomous algorithms to detect zero day phishing websites. Our study has access to the academic feed. Open Phish is used by the antivirus companies Virus Total and Strong Arm. The Open Phish blacklist can be downloaded as a JSON file which typically contains around 5,000 unique URLs.

\paragraph{Phish Tank:} Phish Tank launched in October 2006 and provides a community-based phishing website reporting and verification system. Users of the website can submit URLs of suspected phishing websites; the Phish Tank community then vote as to whether these URLs are phishing or not. Phish Tank is used by the web browser Opera, online reputation and internet safety service web browser plugin Web Of Trust, email provider Yahoo! Mail, and antivirus providers McAfee and Kaspersky~\cite{friendsphishtank}. The Phish Tank blacklist of approved phishing URLs can be downloaded as a JSON file and typically contains around 23,000 unique URLs.

\subsection{Related Work}
Existing literature has explored the effectiveness of malware blacklists \cite{kuhrer2014paint, kuhrer2012empirical} and also phishing attacks in areas such as why they work \cite{dhamija2006phishing}, the effectiveness of toolbars in protecting users \cite{wu2006security, zhang2006phinding}, detection of phishing websites \cite{zhang2007cantina}, the effectiveness of web browser warnings \cite{egelman2008you}, demographic analysis of phishing susceptibility and effectiveness of interventions \cite{sheng2010falls}, and a study to determine a baseline for phishing campaign success \cite{jagatic2007social}. There are also various techniques to prevent phishing attacks including  Dynamic Security Skins \cite{dhamija2005battle}, Trusted Devices \cite{parno2006phoolproof} along with educational aspects of phishing training including PhishGuru \cite{kumaraguru2009phishguru} and the game Anti-Phishing Phil \cite{sheng2007anti}; the effectiveness of these two educational approaches were analysed \cite{kumaraguru2010teaching}. Previous studies have also developed techniques to detect spam, phishing and malware on Twitter, such as looking at redirection chains to detect suspicious URLs \cite{lee2012warningbird}, analysing suspended accounts \cite{thomas2011suspended}, and using social graph models  \cite{wang2010don}. Phishers that use URL shortening services to masquerade phishing URLs on Twitter have also been studied \cite{chhabra2011phi}. 

Two key studies, carried out in 2007~\cite{ludl2007effectiveness} and 2009~\cite{sheng2009empirical} focused on phishing blacklists and how effective they are at protecting users from phishing email attacks, paying particular attention to the delay from an email containing a phishing URL being received to that URL appearing in a blacklist. We focus on Twitter as a delivery platform rather than e-mail.

Whilst these previous studies have looked at the phishing landscape in terms of detecting and preventing phishing attacks, they have not focused specifically on the relationship between blacklists and phishing and malware attacks on Twitter. However, in a 2010 study, Grier \textit{et al.}~\cite{grier2010spam} characterised phishing, malware and scam URLs posted to Twitter. As part of their broad study, of which their overall aim was to characterise spam on Twitter, they analysed blacklist performance, looking at the blacklists GSB, Joewein, and URIBL. One of their main findings was that malicious URLs either appeared in the GSB blacklist, on average, 29.58 days before being tweeted or, if the URLs were not blacklisted at time of tweeting, it took, on average, 24.9 days for the GSB blacklist to detect the URLs. Phishing URLs either appeared in the GSB blacklist, on average 2.57 days before being tweeted, or, if not in the blacklist at time of tweet, an average of 9.01 days after tweeting. 

Whilst the Grier \textit{et al.} study looked at the delay for tweeted URLs to appear in a blacklist, it treated multiple tweets of the same URL as being unique, independent events. We take a different approach. We focus on the delay time between when a blacklisted URL is \textit{\textbf{first}} tweeted to when it first appears in a blacklist such as GSB. We believe this provides a more accurate measurement to ascertain the effectiveness of Twitter URL blacklisting. This is because it enables us to determine how long users are exposed to a specific attack URL since it was first posted to Twitter. One of the main problems with the methodology in~\cite{grier2010spam} is that a URL may be tweeted at a certain point in time, then tweeted again on multiple occasions at much later dates, closer to the point at which that URL becomes blacklisted. This then skews the results because the average delay time for that URL to become blacklisted, when calculated using all tweet times containing that URL, will appear to be smaller than the time of first tweet to blacklist delay. This will tend to underestimate the exposure of users.

A missing detail from~\cite{grier2010spam} is how the historical blacklist data from GSB was obtained. Our study uses timestamps of when URL hash prefixes were downloaded into our local copy of GSB to determine when a URL first appeared in the GSB blacklist. Grier \textit{et al.}~\cite{grier2010spam} were also not specific about which version of Twitter's Stream API they use, other than mentioning that it is a 10\% feed. It is important to note that a 10\% feed in 2010 will have produced approximately 3.5 million tweets per day -- similar to the 3 million tweets per day that we collect in our study.  The methodology section of our paper explains what version of Twitter's Stream API we used, in an effort to improve the reproducibility of our study. 

It is important to note that the aim of~\cite{grier2010spam} was to characterise spam on Twitter, looking at phishing, malware, and scams; that study touched on blacklist performance as part of an overall, broad analysis of spam on Twitter. In contrast, our paper aims to assess how effective Twitter's use of blacklists are at protecting its users from phishing and malware attacks. In contrast, we carry out a more fine-grained and in-depth study into the effectiveness of blacklists on Twitter, particularly focusing on delay periods. As well as replicating the relevant experiments from~\cite{grier2010spam}, we also introduce a new methodology to measure the delay between when a blacklisted URL is \textbf{\textit{first}} tweeted to when it first appears in a blacklist. We also add the Phish Tank and Open Phish phishing blacklists to our study. Finally, and importantly, we check redirection chains for each tweeted URL, since blacklisted URLs may be hidden in such chains.


\section{Design}
\label{sec:design}


\subsection{Overview}

\begin{figure}[t!]
\centering
\includegraphics[width=0.4\textwidth,trim={2.5cm 3.5cm 5cm 2cm},clip]{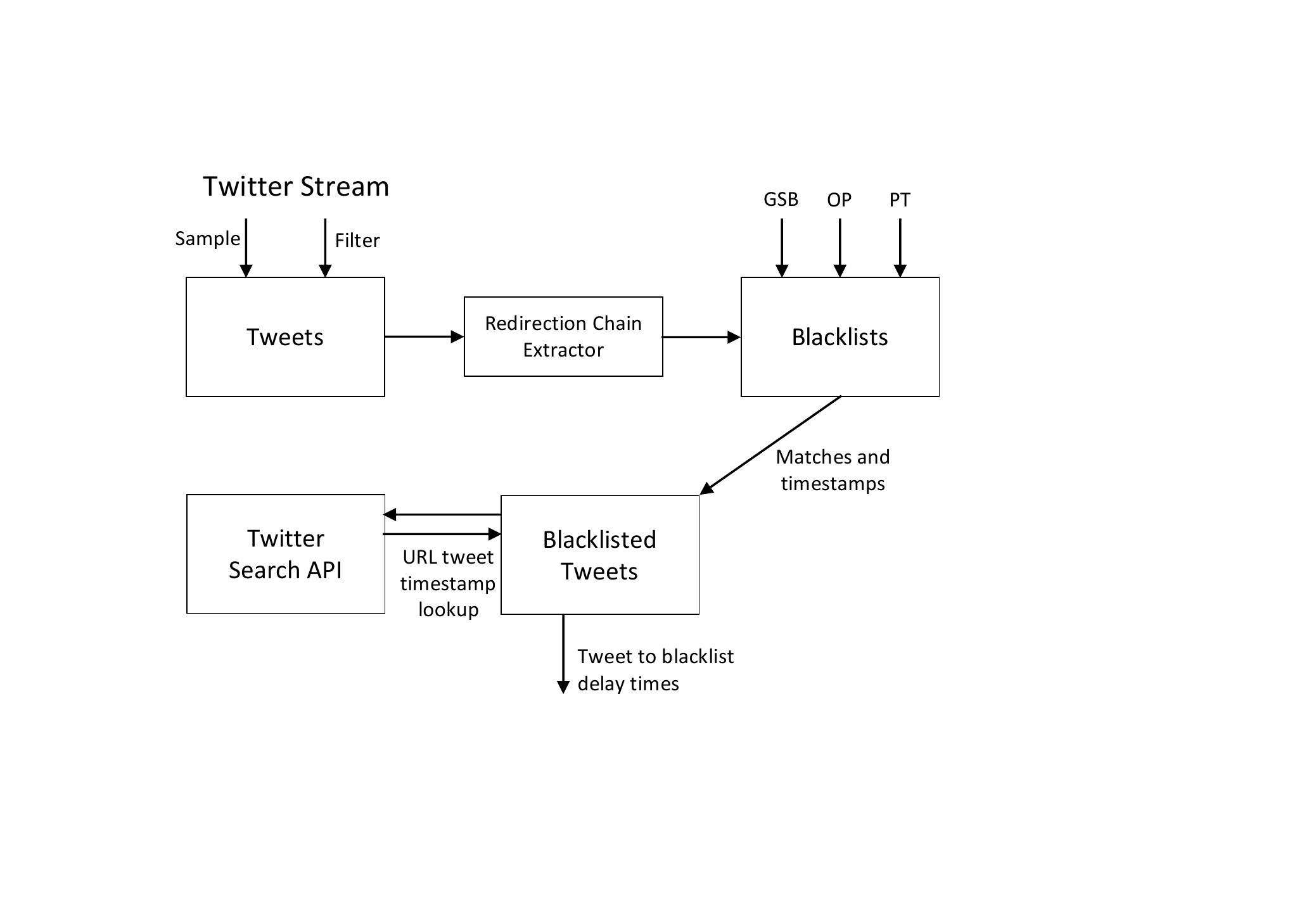}
\caption{Infrastructure design architecture.}
\label{design-architecture}
\end{figure}

The infrastructure used to carry out experiments for our study consists of a tweet collection system that receives both sample tweets and also tweets containing URLs from Twitter; a database to store these collected tweets; a URL redirection chain extractor; a blacklist system to store, update and perform lookups against 3 popular blacklists; a database to store tweeted URLs which have appeared in a blacklist; a Twitter search API lookup system to determine when tweeted URLs first appeared on Twitter; and a measurement system to calculate delays from URLs being tweeted to appearing in blacklists. The overall architecture of this system can be seen in Figure \ref{design-architecture}.

\subsection{Data Collection}

The first requirement for our study is a source of live tweets from Twitter. To achieve this we setup two sources of incoming tweet feeds using Twitter's Stream API. The first stream is, approximately, a 1\% sample of all global tweets and the second stream is, approximately, a 1\% sample of all global tweets that contain one or more URLs. The first stream is used to provide a general picture of Twitter activity during collection and the second stream is used to carry out our blacklist delay analysis by looking up the tweeted URLs in various blacklists. Both of these tweet streams are saved locally in a database.

The second requirement for our study is a way to store and search various blacklists. For our study 3 blacklists are used: GSB, Open Phish, and Phish Tank. The GSB blacklist includes both social engineering and malicious URLs. Our system regularly obtains the latest copies of these blacklists and saves them locally in a database on our system. Tweeted URLs from our collection of tweets can then be searched for in these 3 blacklists. 

\subsection{Methodology}
As described in the previous subsection, tweets containing URLs are collected from the Twitter stream and saved into a local database. 3 blacklists are also stored in the local database. In order to determine which tweeted URLs appear in the 3 blacklists two systems are used: fast and slow. The fast system checks the 3 million most recently tweeted URLs (equivalent to about 24 hours of tweets) against the GSB blacklist every 10 minutes and the Open Phish and Phish Tank blacklists every hour. We determined this 10 minute update frequency by carrying out a small-scale study to observe how frequently the GSB Update API blacklist updates. Open Phish and Phish Tank refresh their blacklists every 60 minutes. The slow system checks all tweeted URLs we have collected since our experiment began and performs a lookup on the latest versions of all 3 blacklists. This slow lookup system will complete its cycle of all URLs relatively quickly at first but increase in duration as the number of URLs in the experiment grow. The main slowdown in this lookup system is that the GSB API requires any hash prefix match to be sent to GSB's servers for the full hash to be downloaded then checked for a match. This system is necessarily slower in its operation, taking a number of hours to complete a pass over our full collection of tweets. The reason for these two lookup systems (fast and slow) is because GSB does not include a ``time of inclusion'' for blacklisted URLs. This system helps us to determine when URLs appear in the GSB blacklist, with finer resolution on URLs that are tweeted within 24 hours. The outcome is that we can produce more accurate results in our measurements.

During the experiment it was discovered that the library implementation we use for GSB's Update API also stores timestamps for when blacklisted URL hash prefixes were added to the local database. We then built a system to lookup each blacklisted URL's hash prefix timestamp to determine when each URL was added to our local copy of the blacklist. The GSB Update API library stores each blacklisted URL as a 4-byte SHA256 hash prefix; due to the small size of these URL hash prefixes, there is a chance that collisions may occur. Because of this, only hash prefix lookups that had zero collisions were used for the experimental results. This additional system complements the previously mentioned fast and slow lookup systems because the new system will produce more accurate results for when a URL is already in GSB -- particularly if a URL has been in GSB for a significant amount of time. The fast and slow systems are still required for when tweeted URLs are not in GSB at time of tweet.

For a tweeted URL, there could be a number of hops or redirections that are made before arriving at the final landing page. For this reason a redirection chain extractor is used to check each URL contained within in a redirection chain against each of the blacklists. The technicalities of this redirection chain extractor system are explained in more detail in the implementation section.

When calculating the time from a tweet appearing in the Twitter Stream feed to appearing in one of the 3 defined blacklists, some tweeted URLs may have previously appeared on Twitter prior to being received in the Twitter Stream feed. To compensate for this, we carry out another experiment. In this experiment, when computing delays from time of tweet to time of blacklist appearance, we built a system to lookup each blacklisted URL in Twitter's Search API. Our system can determine when the URL was first tweeted; this timestamp can then be used to calculate the delay between first tweet and first blacklist appearance, therefore increasing the accuracy of the measurement. Limitations of using this approach, as stated in Twitter's Search API documentation, are that it is limited to 7-10 days, it is not an exhaustive source of tweet. Therefore not all tweets will be indexed or made available via the search interface. 

\subsection{Overview of Experiments}

Our first experiment analyses tweets collected from Twitter's Stream API with the sample method; these sample tweets are collected during the same time frame as the URL-containing tweets. This experiment shows us the ratio of URL containing to non-URL containing tweets along with a breakdown of the numbers of tweets received per day. 

Our second experiment replicates one of the experiments carried out by Grier \textit{et al.}~\cite{grier2010spam} in which the delay from a URL being tweeted to appearing in the blacklists is calculated. This experiment shows what has changed since the 2010 study -- particularly since Twitter's active user base has grown from 30 million in 2010 to over 330 million in 2017 and total number of daily tweets has grown from 35 million in 2010 to over 500 million in 2017.

Our third experiment uses a different methodology to the 2010 study \cite{grier2010spam}, in that we use the timestamp for when a blacklisted URL was \textbf{\textit{first}} tweeted to calculate delay to first appearing in a blacklist. If a URL is tweeted multiple times then only the first tweet to contain that URL will be used to calculate delay. This measurement is important as it allows us to determine how long it takes for URLs to appear in blacklists after they are first tweeted. This experiment also includes the Phish Tank and Open Phish databases. 

Our fourth experiment is an improvement on the previous experiment in that the Twitter Search API system is used to determine when a URL was first tweeted. Within Twitter's Search API limit, of 180 calls per 15 minute window, URLs that appear in blacklists are searched for on Twitter to determine their original tweet date. This allows us to determine, with more accuracy than the previous experiment, when a URL was first tweeted (i.e. if we did not receive the original tweet containing a given URL in our Twitter Stream). This also provides us with the worst case scenario measurement. 

Our fifth experiment analyses for how long blacklisted tweeted URLs remain in the GSB blacklist for. In order to carry out this experiment the timestamp for when a URL first appears in the Twitter Filter (URL) Stream is compared against the last time the system matched the same URL in the GSB blacklist. The difference between these two timestamps is used as the measurement.

\section{Implementation}
\label{sec:implementation}

Our entire system is implemented on a virtual machine running the Ubuntu operating system, version 16.04 LTS, 8 core CPU, 24 GB RAM. The measurement framework is written in the programming language Python. 

Our Twitter collection system uses Twitter's Stream API, implemented via the Tweepy~\cite{tweepy} library. After authorising \textit{Tweepy} to access Twitter, the \textit{sample()} and \textit{filter()} methods are used to collect sample and URL containing tweets. The filter method uses keywords \textit{``http''} and \textit{``https''} to filter out tweets containing URLs. All data received from Twitter's Stream API, using these two methods, is stored in a MySQL~\cite{mysql} version 5.7.19-0ubuntu0.16.04.1 database in two tables for sample tweets and URL-containing tweets, respectively.

The URL redirection chain extraction system uses Python's \textit{Requests} library~\cite{python-requests} to send a HTTP request for each URL using a Macintosh Safari user agent header so the request appears to come from a regular user via the Safari web browser.  The reason for setting this header is so the request extracts the same redirection chain that a legitimate user would see and not a redirection chain that a bot would see -- therefore reducing bias in our results. The \textit{Request} library's \textit{Response} object contains a \textit{History} property which consists of a list of \textit{Response} objects that were created to complete the HTTP request. This list is then used to extract the redirection chain for a given URL in our system.

We use 3 blacklists in our system: GSB, Open Phish and Phish Tank. To implement our GSB lookup system, the library \textit{gglsbl}~\cite{gglsbl} version 1.0.0 is used. This library allows our system to fetch the latest GSB hash prefixes and also perform lookups against the database. The library uses the SQLite~\cite{sqlite} database for storing GSB data. The library contains a method \textit{update\_hash\_prefix\_cache()} which is used to update the URL hash prefix database. This method is called every 10 minutes in the fast GSB lookup system and at the beginning of each cycle of the slow GSB lookup system.

An important modification was made to the \textit{gglsbl} library to improve lookup times for large numbers of URLs. The method \textit{lookup\_url()} is used to lookup an individual URL in the local hash prefix database. It does this by performing an SQLite search for that URL's hash prefix. This lookup technique caused a bottleneck when testing the system on large volumes of URLs, therefore the library was modified to output a Python dictionary (hash table) of all URL hash prefixes. Our system can then perform a lookup for each tweeted URL's hash prefix against this dictionary. Since the Python dictionary implementation uses a hash map, the typical time complexity for this lookup is constant; $O(1)$. This means lookups are considerably faster than using the off-the-shelf version of the GSB library. 

During our experiment, we observe that the GSB blacklist typically contains approximately 4.8 million URL hash prefixes of which approximately 3.1 million are unique. Of these, there are approximately 1 million unique URL hash prefixes labelled malware and approximately 1.8 million unique URL hash prefixes labelled social engineering. The remaining URL hash prefixes labels are not used in our study.

Both the Phish Tank and Open Phish datasets are download as JSON files from their websites. The URL entries from these files are then extracted and saved into our local MySQL database. Metadata stored along with URLs includes discovery timestamps from the blacklists and timestamps for when URLs were added to our database. Both datasets are downloaded every hour and new entries saved in the local database. URL lookups against these two databases are completed by importing all URLs from both databases and storing them in a Python dictionary in order to perform faster lookups, as per our GSB lookup implementation.

Our Twitter Search API lookup system uses the \textit{Tweepy} library to interact with Twitter's Search API. After authorising \textit{Tweepy} to access Twitter, the \textit{Search} method is used to search for a given URL. This method will return the oldest tweet in Twitter's search history, that contains a given URL string, if it can be found.

\section{Results}
\label{sec:results}


\subsection{Twitter Dataset Analysis}

\begin{table}[t!]
\begin{center}
\begin{tabular}{ | l | r | r | } 
  \hline
  & October & November \\ 
  \hline
  Twitter Sample & 105,306,234 & 100,817,746 \\  
  \hline
  | URL & 24,085,266 & 23,478,257 \\ 
  \hline
  | Non-URL & 81,220,968 & 77,339,489 \\
  \hline
  Twitter Filter (URL) & 91,871,659 & 90,719,779 \\ 
  \hline
\end{tabular}
\end{center}
\vspace{0.1cm}
\caption{Total number of collected Twitter sample and Twitter filter (URL) stream tweets, October and November 2017.}
\label{total-sample-tweets}
\end{table}

\begin{figure}[t!]
\centering
\includegraphics[width=0.5\textwidth]{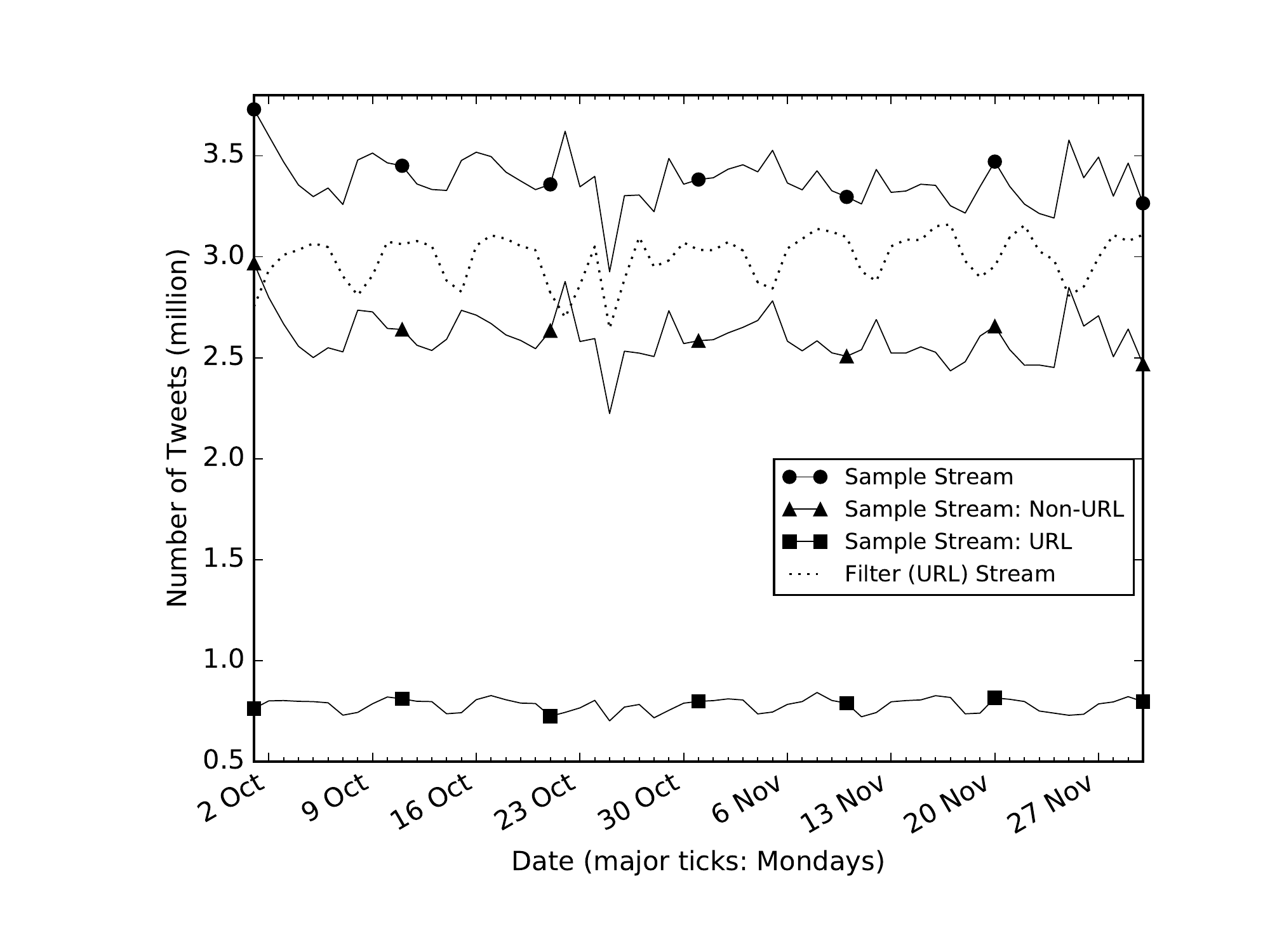}
\caption{Total tweets collected per day: sample stream \& filter (URL) stream API, October and November 2017.}
\label{tweets-per-day-oct-nov-2017}
\end{figure}

\begin{figure}[htp] 
\centering
    \subfloat[Social Engineering, Oct \& Nov 2017]{
        \includegraphics[width=0.25\textwidth]{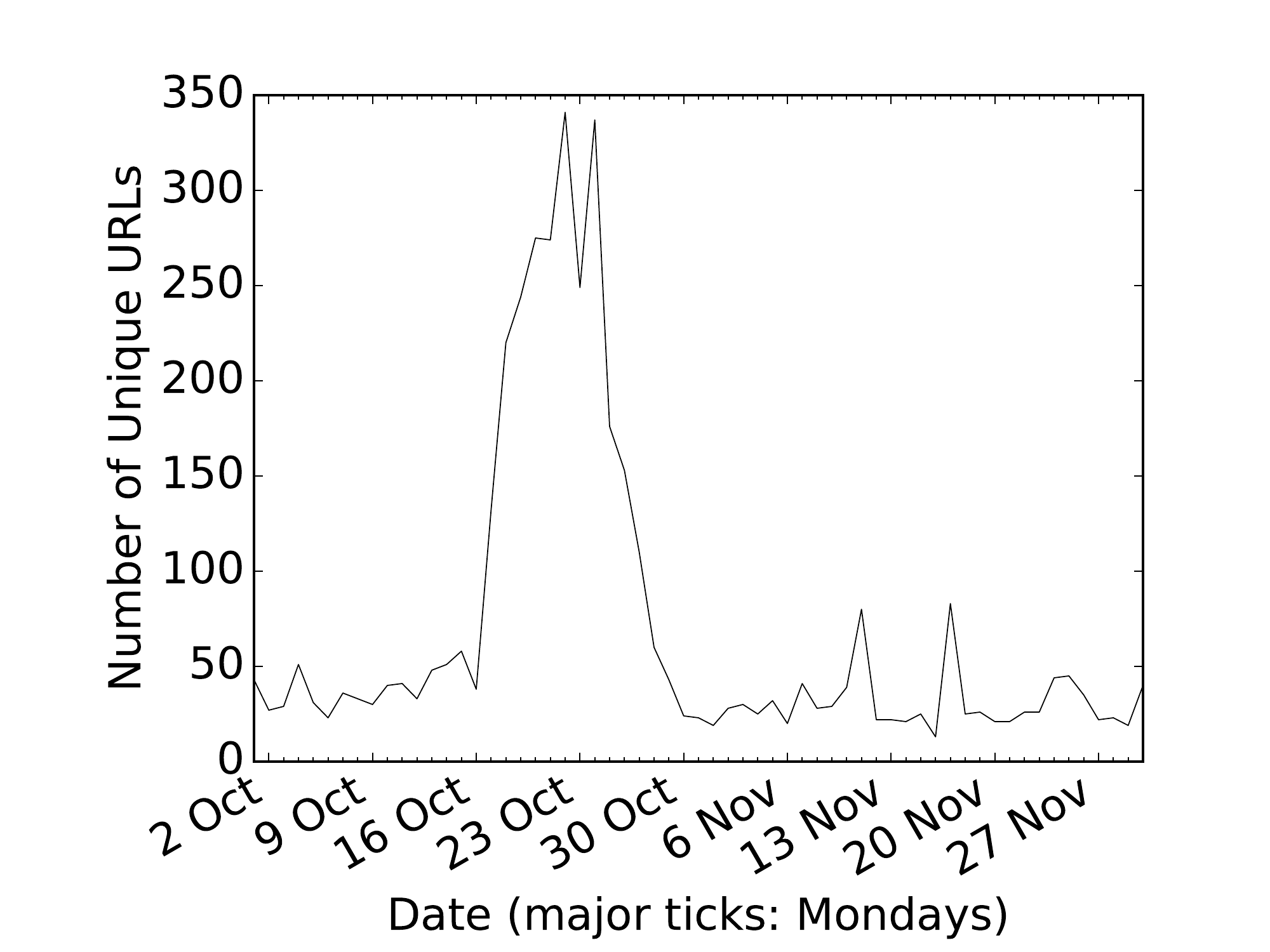}
        \label{fig:phishing-urls-per-day-oct-nov}
        }
    \subfloat[Malware, Oct \& Nov 2017]{
        \includegraphics[width=0.25\textwidth]{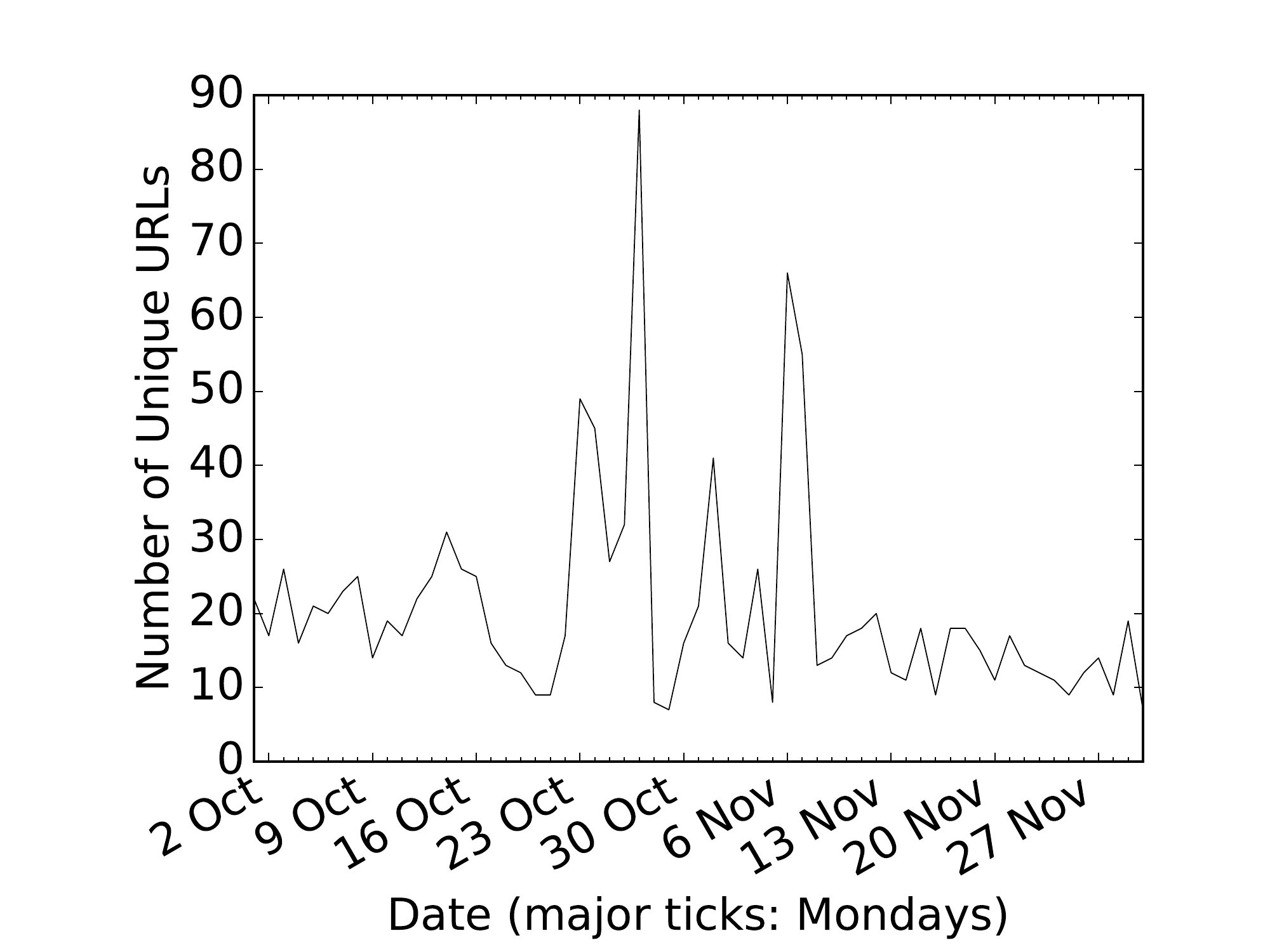}
        \label{fig:malware-urls-per-day-oct-nov}
        }
        \caption{Total unique first tweeted social engineering \& malware URLs per day that first appeared in GSB blacklist within 1 month before or after tweet in October \& November 2017.}
\end{figure}

This first results subsection provides an overview of the dataset
obtained from collecting tweets via the Twitter Stream API
\textit{Sample} method during October and November 2017. We collect,
approximately, 3.4M sample tweets and 3M URL-containing
tweets per-day.
Overall, the Twitter Stream Sample collected 105,306,234 tweets in October 2017
and 100,817,746 tweets in November 2017; of these, only 23\% contained
URLs. The Twitter Stream Filter (URL) collected 91,871,659 tweets in
October 2017 and 90,719,779 in November 2017, as shown in Table
\ref{total-sample-tweets}. Figure \ref{tweets-per-day-oct-nov-2017}
shows the per-day total number of Twitter Sample Stream tweets,
including URL and non-URL containing tweets, along with total number
of Twitter Filter (URL) Stream tweets collected in October and November 2017.

There are 10,029 unique URLs that first appeared in the Twitter Stream Filter (URL) in either October or November that subsequently appeared in one of the GSB, Open Phish or Phish Tank blacklists at some point during our experiments. Of these URLs, 5,464 appeared in one of the blacklists within 1 month before or after first appearing in the Twitter Filter (URL) Stream, as seen in Table \ref{tab:collected-blacklisted-urls}. It is interesting to note that only 9 URLs from Open Phish and Phish Tank appeared in the Twitter Filter (URL) stream during the October and November timeframe. In October, of the 2 Open Phish URLs that were tweeted, 1 had been added to the Open Phish blacklist on the 22nd August 2017 and the other had a delay of 12 days from date first tweeted to appearing in the blacklist. Of the 2 Phish Tank URLs from October: one had been tweeted on the 15th October 2017, but was blacklisted by Open Phish on 1st September 2017, the other was blacklisted approximately 5 minutes after being tweeted. For November: the 1 Open Phish URL appeared in the blacklist approximately 5 minutes after being tweeted. For the 4 Phish Tank URLs, blacklist delays were approximately 32 minutes, 35 minutes, 21 days and 9 days after tweet. Considerably fewer tweeted URLs appeared in the Open Phish and Phish Tank blacklists compared to GSB. One reason for this difference may be that the GSB blacklist contains approximately 3 million URLs whereas the Phish Tank and Open Phish blacklists contain 28,000 URLs combined; there are fewer URLs for Phish Tank and Open Phish to detect. Another possibility is that Twitter is using the Phish Tank and Open Phish blacklists and therefore preventing users from tweeting URLs contained within these blacklists. However, if that were the case, then we would still see URLs in the Twitter Stream before they appear in the Open Phish or Phish Tank blacklists. 

Figures \ref{fig:phishing-urls-per-day-oct-nov} and \ref{fig:malware-urls-per-day-oct-nov} show the total number of unique URLs per day that first appeared in the Twitter Filter (URL) Stream in the given month that subsequently first appeared in the GSB Blacklist, as either social engineering or malware, within 1 month before or after appearing in the Twitter Filter (URL) Stream for October and November 2017. 

\paragraph{\textbf{Findings:}} These results show that we collected,
approximately, 3.4M sample and 3M URL-containing tweets per day
throughout October and November 2017. Of these, 5,464 unique URLs
appeared in one of the 3 blacklists within 1 month before or after
first appearing in the Twitter Stream Filter (URL). This volume of tweets provides us with a good amount of data to explore delay times, click metrics, and overall time in GSB in the upcoming sections. We also see there are only 9 URLs from the Open Phish and Phish Tank blacklists. This may possibly be because the Phish Tank and Open Phish blacklists contain fewer URLs (approximately 28,000) compared to GSB (approximately 3 million).

\begin{table}[t!]
\small
\begin{center}
\begin{tabular}{ | l | r | r | r | r | } 
 \hline
   & \multicolumn{2}{|c|}{October} & \multicolumn{2}{|c|}{November} \\ 
  \hline
  Blacklist & URLs & Domains & URLs & Domains \\ 
  \hline
  GSB SE$^{\star}$ & 4,912 & 397 & 2,495 & 268 \\  
  \hline
  GSB SE$^{\dagger}$ & 3,273 & 212 & 930 & 182 \\    
  \hline
  GSB SE$^{\mathsection}$ & 295 & 89 & 294 & 73 \\ 
  \hline
  GSB Malware$^{\star}$ & 1,563 & 250 & 1,054 & 144 \\ 
  \hline
  GSB Malware$^{\dagger}$ & 718 & 82 & 543 & 65 \\ 
  \hline
  GSB Malware$^{\mathsection}$ & 230 & 37 & 131 & 29 \\
  \hline
  Open Phish$^{\star}$ & 2 & 2 & 1 & 1 \\
  \hline
  Open Phish$^{\dagger}$ & 1 & 1 & 1 & 1 \\
  \hline
  Phish Tank$^{\star}$ & 2 & 2 & 4 & 3 \\
  \hline
  Phish Tank$^{\dagger}$ & 1 & 1 & 4 & 3 \\
  \hline
\end{tabular}
\end{center}
\vspace{0.1cm}
\caption{Number of unique, blacklisted social engineering (SE) and malware URLs \& domains first tweeted in October and November 2017.
\\$^{\star}$Blacklisted anytime during experiment.
\\$^{\dagger}$Blacklisted within 1 month from first tweet date.
\\$^{\mathsection}$Blacklisted within 1 month from first tweet date and using Twitter's Search API to determine URL first tweet date.}
\label{tab:collected-blacklisted-urls}
\end{table}

\subsection{Blacklist Delays -- All Blacklisted Tweets}
In this subsection we replicate one of the experiments carried out in 2010 by Grier \textit{et al.} \cite{grier2010spam}. It is important to note that it is difficult to replicate their study exactly because their methodology is not completely explained in their paper. Specifically, they do not explain how a historical copy of the GSB blacklist is acquired or if they allow a delay period of 1 month before and after every URL is tweeted. To the best of our knowledge, based on their paper, this is a replication of one of the experiments in their study. 

In this experiment the delay period for a tweeted URL to appear in the GSB blacklist is calculated using time of tweet to time first appearing in blacklist. If a URL is tweeted multiple times then each posting is treated as a unique, independent event. This is the same methodology used by \cite{grier2010spam}. In our results a negative delay value represents a URL that appears in the blacklist before it is tweeted and a positive delay value represents a URL that appears in the blacklist after being tweeted. This is because we are measuring the delay from a URL being tweeted to first appearing in a blacklist, so a delay value of 20 days means it took 20 days from that URL being tweeted to appearing in a blacklist. The Grier \textit{et al}.~\cite{grier2010spam} study uses lead and lag times in their measurements, where a lead time signifies a URL that appears on Twitter before being blacklisted and a lag time is used to denote a URL that appears in a blacklist after being tweeted. As a result, their lead times are positive and lag times are negative values.

\begin{figure*}[htp] 
    \subfloat[Social Engineering Oct 2017]{
        \includegraphics[width=0.25\textwidth]{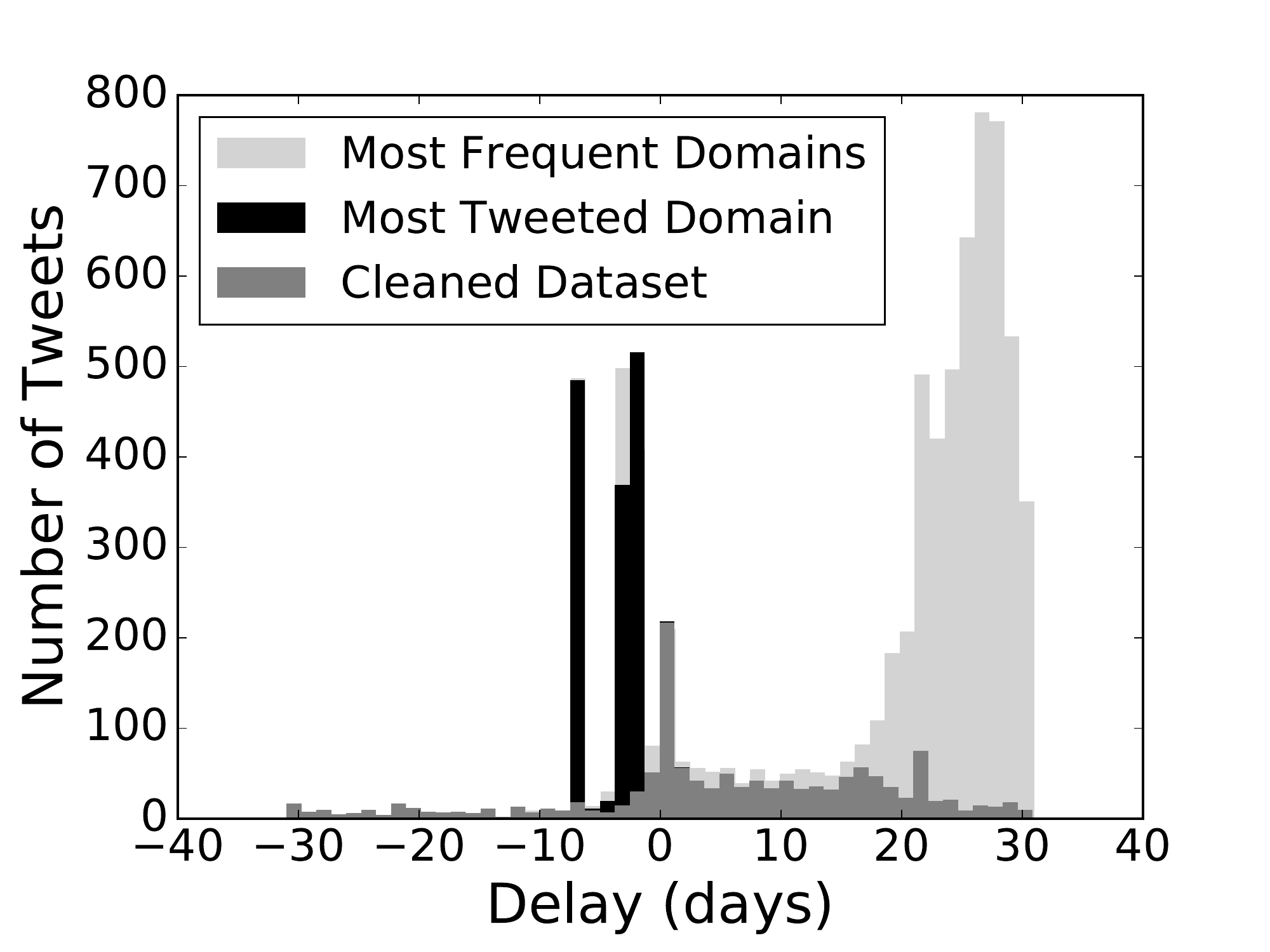}
        \label{fig:phishing-tweets-delay-histogram-oct-2017-top-domains-removed}
        }
    \subfloat[Social Engineering Nov 2017]{
        \includegraphics[width=0.25\textwidth]{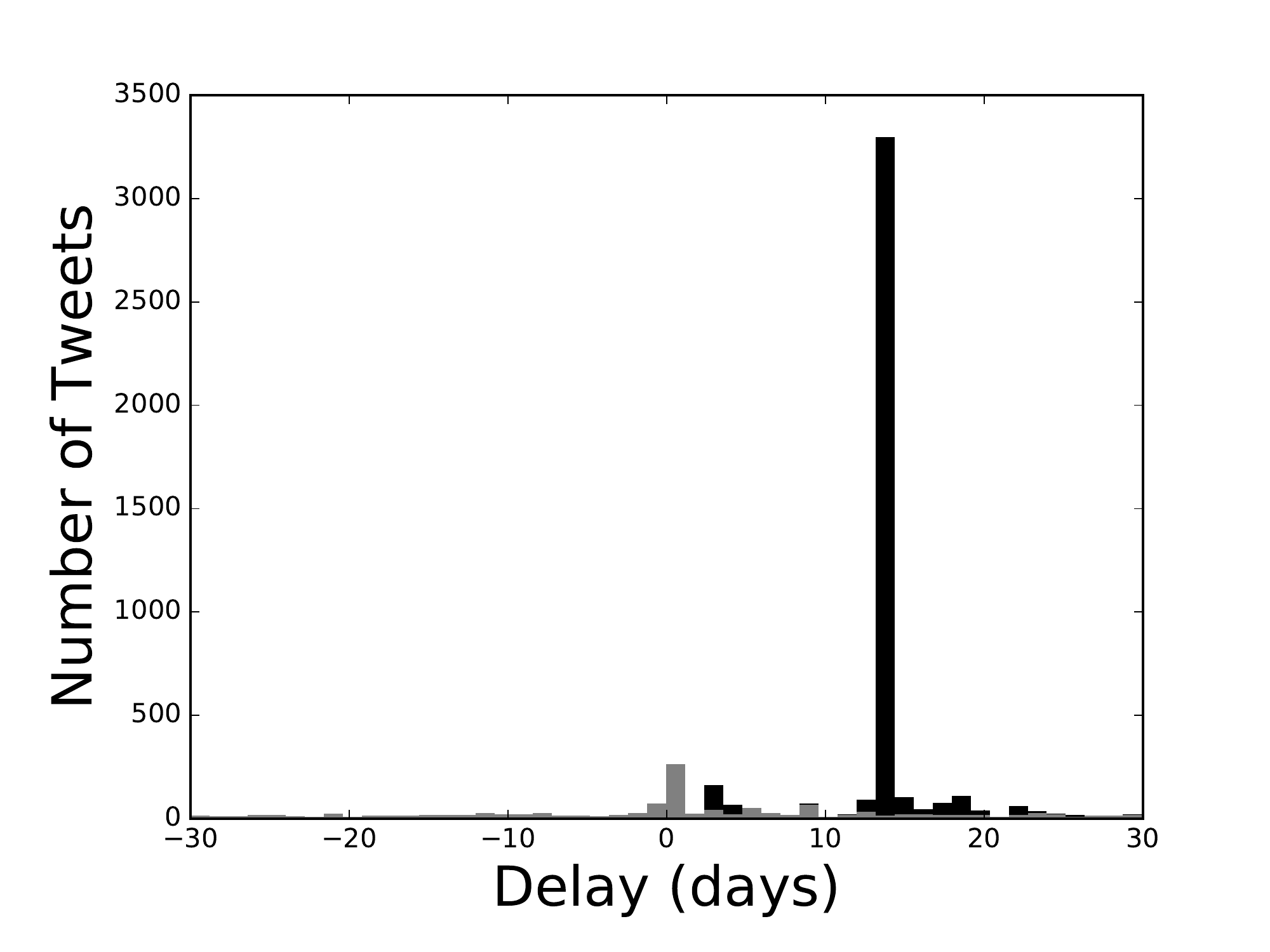}
        \label{fig:phishing-tweets-delay-histogram-nov-2017}
        }
    \subfloat[Malware Oct 2017]{
        \includegraphics[width=0.25\textwidth]{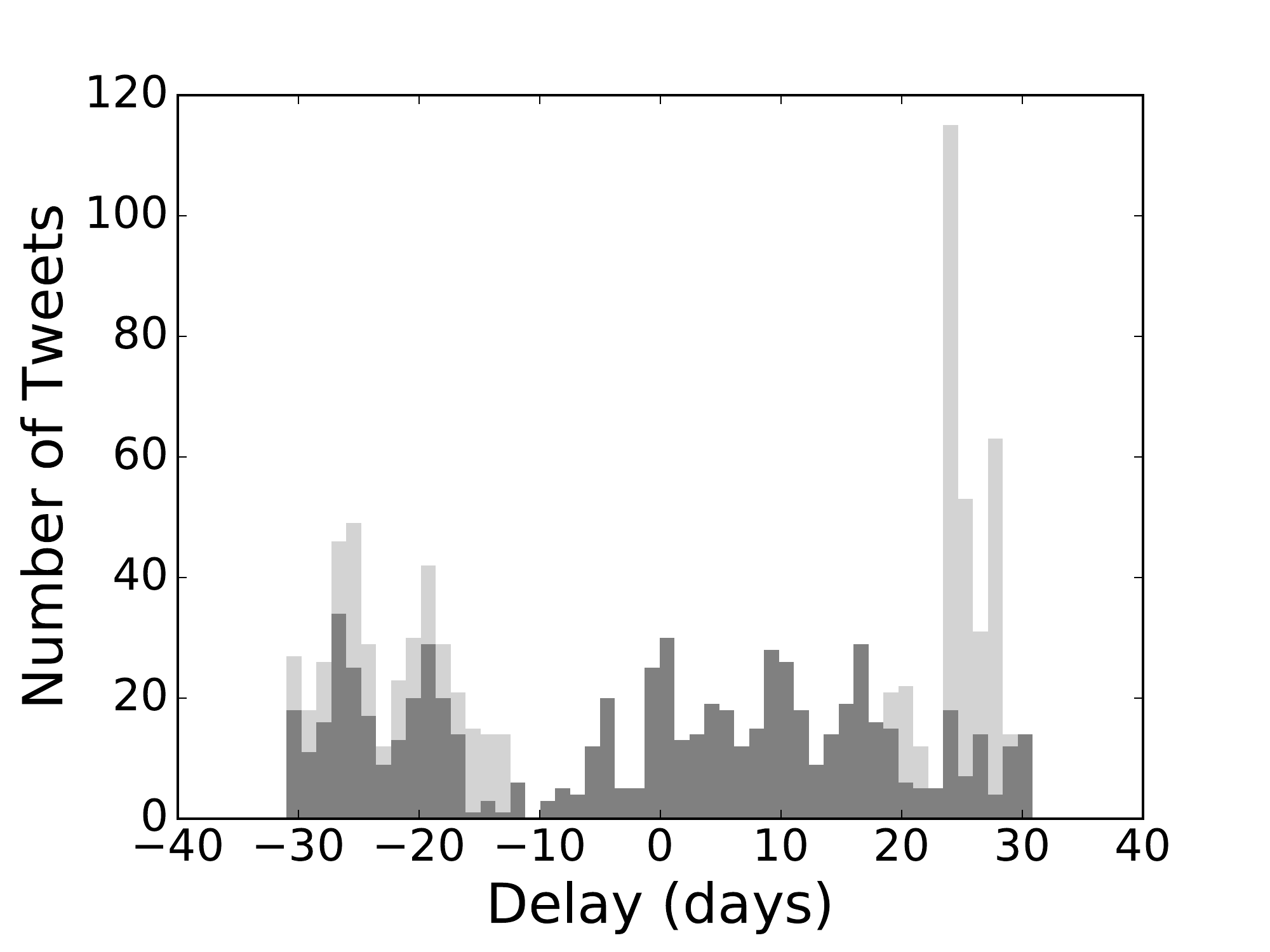}
        \label{fig:malware-tweets-delay-histogram-oct-2017-top-domains-removed}
        }
    \subfloat[Malware Nov 2017]{
        \includegraphics[width=0.25\textwidth]{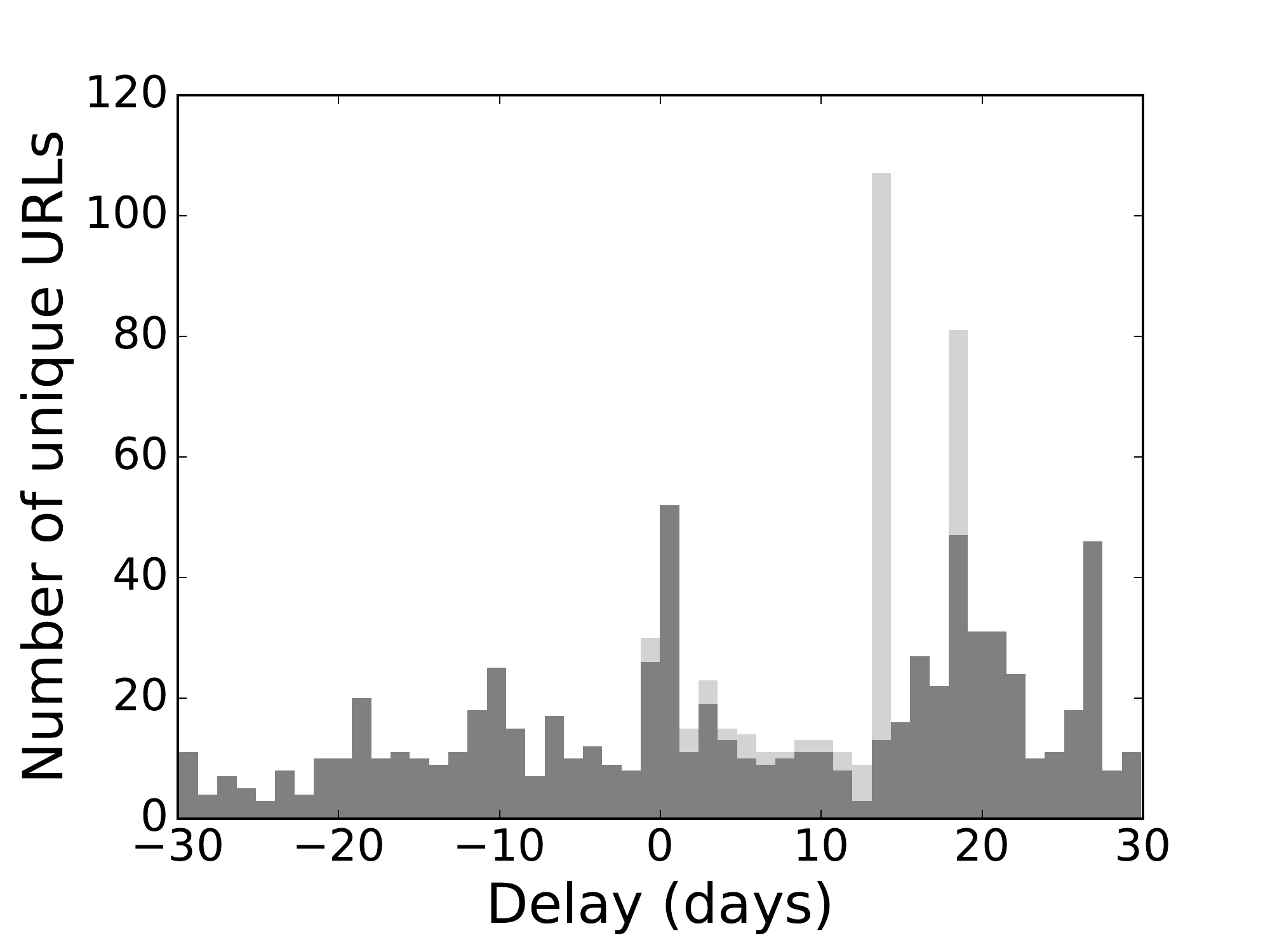}
        \label{fig:malware-tweets-delay-histogram-nov-2017-minus-top-domains}
        }
        \caption{Delay time for all tweets containing GSB blacklisted URLs (including most frequent domain names) labelled social engineering and malware, November and October 2017.}
\end{figure*}

Our first experiment looks at URLs that were tweeted during October and November 2017 which were subsequently labelled as social engineering in the GSB blacklist within 1 month before or after being tweeted. We believe this is the most accurate way to carry out this measurement since the same timeframe is applied to all individual tweets, regardless of when they were tweeted in the month. This methodology is not defined in \cite{grier2010spam} so it may affect the comparison. Timestamps for when tweets are received from the Twitter Filter (URL) Stream are used as tweet date and URL hash prefix timestamps from the GSB blacklist library are used to determine time first appeared in GSB blacklist to calculate total delay from tweet to blacklist, as described in the methodology section of this paper. During this experiment a total of 7,597 tweets containing social engineering URLs in the GSB blacklist were recorded in October and 5,193 in November, as seen in Figures \ref{fig:phishing-tweets-delay-histogram-oct-2017-top-domains-removed} and \ref{fig:phishing-tweets-delay-histogram-nov-2017}. We then carry out the same experiment for malware URLs: a total of 1,110 tweets containing malware URLs embedded the GSB blacklist were recorded in October and 914 in November, as seen in Figures \ref{fig:malware-tweets-delay-histogram-oct-2017-top-domains-removed} and \ref{fig:malware-tweets-delay-histogram-nov-2017-minus-top-domains}. An additional step we take in our experiments, which was not carried out in \cite{grier2010spam}, is to further investigate anomalies in these results and to also clean the results by removing the most frequent domain names. This is explained further in the next subsection. For tweets containing blacklisted social engineering URLs In October there is a spike of tweets at -8 days and again between -2 and -4 days. These spikes are caused by one domain name. For tweets containing blacklisted social engineering URLs in November, there is a peak of 3,316 tweets that have a delay time of between 13 and 14 days, as seen in figure \ref{fig:phishing-tweets-delay-histogram-nov-2017}. This spike is caused by one domain name.

When comparing our results to the 2010 study \cite{grier2010spam} it is important to remember that their study had access to a 10\% Twitter feed of approximately 35 million tweets per day in 2010; our 2017 study collects approximately 3 million URL-containing tweets per day -- comparable numbers. The first noticeable difference is that there are a greater number of overall tweets containing blacklisted social engineering URLs in our study. Whereas \cite{grier2010spam} sees a greater number of tweets containing malware URLs appear in GSB after they have been tweeted. We see significantly more tweets containing blacklisted URLs appearing on Twitter after they have appeared in the GSB blacklist for both social engineering and malware URLs. In our study, for social engineering tweets in November, the delay with the greatest number of tweets is 13.5 days with approximately 3,275 tweets. In \cite{grier2010spam}, the delay with the greatest number of tweets approximately -6 days with approximately 58 tweets. In our study, when looking at  social engineering tweets in October, the delay with the greatest number of tweets is 26 days with 790 tweets. This shows that, in our results, there is a greater volume of social engineering tweets appearing on Twitter. The results in \cite{grier2010spam} show that the average lag time for social engineering tweets is 9.01 days and the average lead period is -2.57 days. For malware tweets the average lag time is 24.90 days and the average lead time is -29.58 days. The results for the average lead and lag times for our experiments can be seen in Table \ref{tab:lead-and-lag-averages}. These figures show that the lag time averages can vary depending on if the most frequent domain names are included in the calculation, as is the case for social engineering and malware tweets in October.

\paragraph{\textbf{Findings:}} One of the most significant differences between our results and those in 2010 \cite{grier2010spam} is that, in our results, there are substantially more URLs being posted to Twitter \textit{\textbf{after}} they appear in the GSB blacklist, compared to \cite{grier2010spam}. This suggests that Twitter has altered its filtering process to allow some URLs blacklisted by GSB to be tweeted or they may have stopped using the GSB blacklist altogether and built their own URL filtering system.
        
\begin{table}[t!]
\small
\begin{center}
\begin{tabular}{ | p{2cm} | r | r | r | r | } 
 \hline
   & \multicolumn{2}{|c|}{GSB SE} & \multicolumn{2}{|c|}{GSB Malware} \\ 
  \hline
   & Oct & Nov & Oct & Nov \\ 
  \hline
  Avg. lag (days) & 22.62 & 12.71 & 17.99 & 15.02 \\  
  \hline
  Avg. lag - top domains removed (days) & 11.05 &  & 13.37 & 15.55 \\  
  \hline
  Avg. lead (days) & -5.39 & -12.18 & -19.41 & -12.57 \\  
  \hline
  Avg. lead - top domains removed (days) & -5.46 & & -18.24 & -12.73 \\  
  \hline
\end{tabular}
\end{center}
\vspace{0.1cm}
\caption{Average delay times for all tweeted blacklisted social engineering (SE) and malware URL delays. Lead and lag times indicate appearing in blacklist before or after being tweeted, respectively.}
\label{tab:lead-and-lag-averages}
\end{table}

\subsection{Blacklist Delays -- From Time of First Tweet}
\label{sec:results-time-of-first-tweet}

In this subsection we use a different methodology to \cite{grier2010spam} in that the timestamp for when a blacklisted URL was \textit{\textbf{first}} tweeted is used to calculate delay to first appearing in a blacklist. This new methodology is important as it allows us to determine how long it takes for URLs to appear in blacklists after they are first tweeted -- therefore calculating how long users are exposed to attacks for. If a URL is tweeted multiple times then only the first tweet to contain that URL will be used to calculate delay. One of the main problems with the measurement carried out in \cite{grier2010spam} is that a URL may be tweeted at a certain point in time, then tweeted again on multiple occasions at a much later point in time; closer to the point at which that URL becomes blacklisted. This then skews the results because, in this example, the average delay time for that URL to become blacklisted, when calculated from all tweet times containing that URL, will be less when compared to just the time of first tweet to blacklist delay.

\begin{figure*}[htp] 
    \subfloat[Social Engineering Oct 2017]{
        \includegraphics[width=0.25\textwidth]{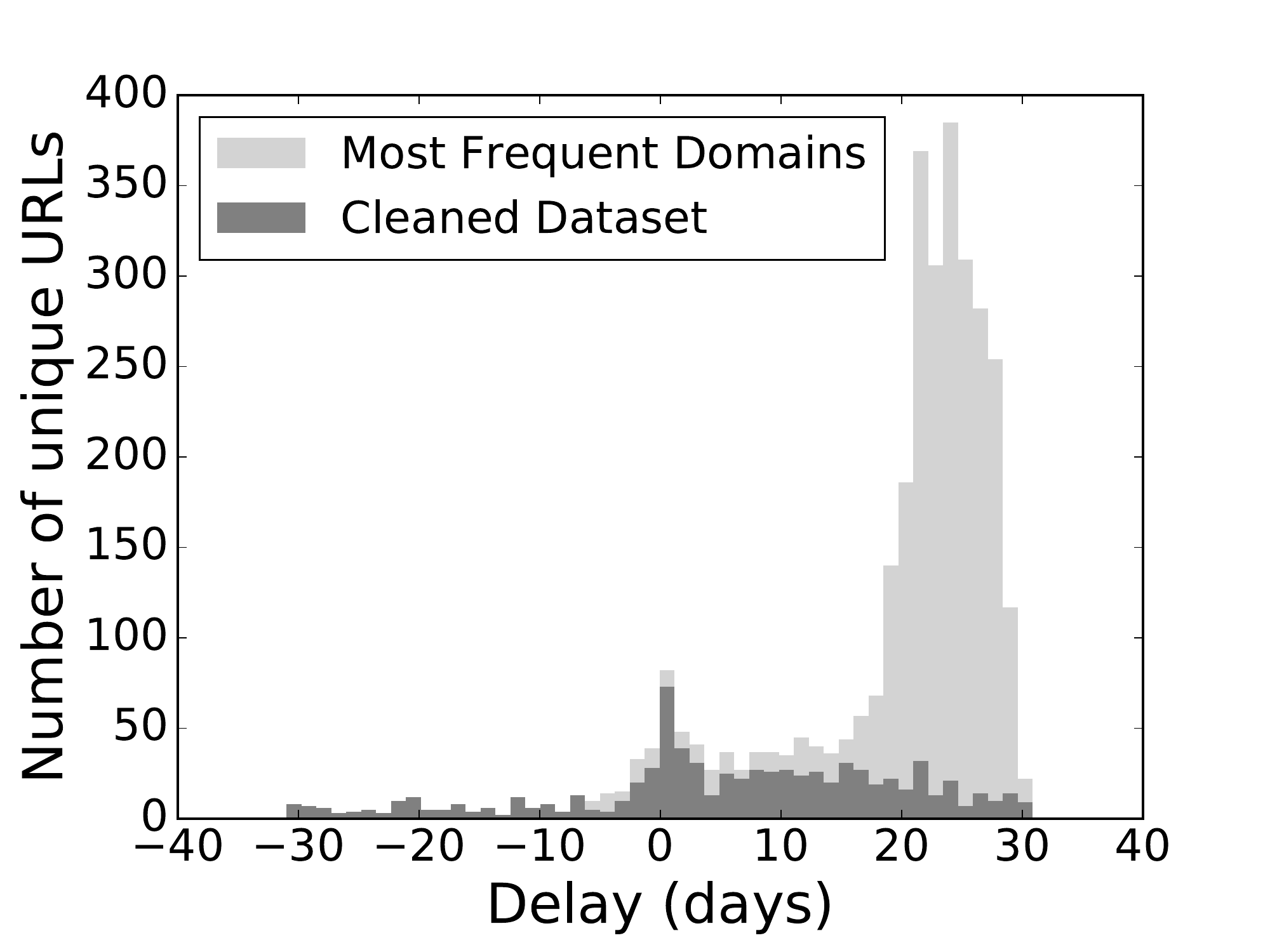}
        \label{fig:phishing-urls-delay-histogram-oct-2017-minus-top-7-domains}
        }
    \subfloat[Social Engineering Nov 2017]{
        \includegraphics[width=0.25\textwidth]{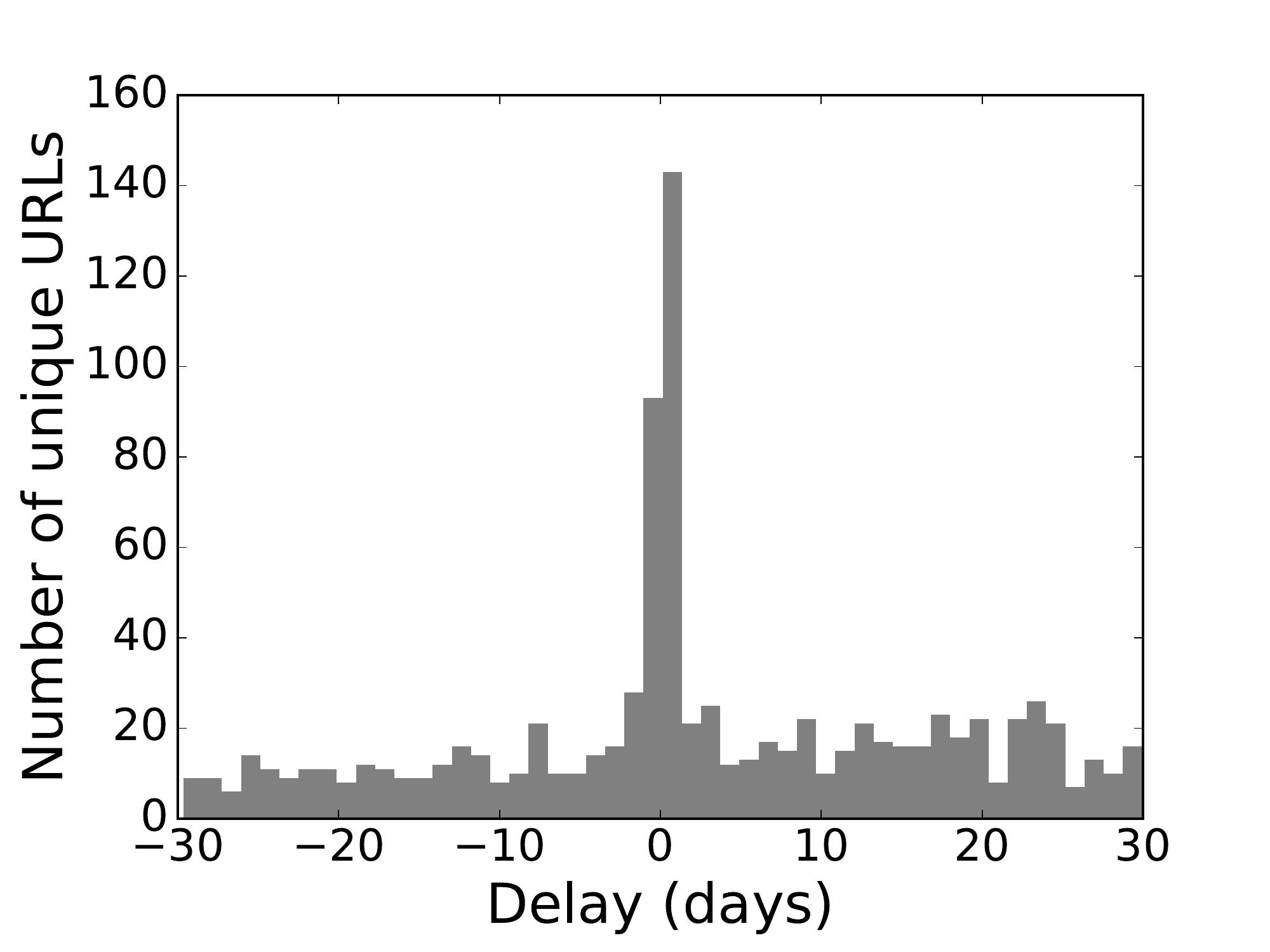}
        \label{fig:phishing-urls-delay-histogram-nov-2017}
        }
    \subfloat[Malware Oct 2017]{
        \includegraphics[width=0.25\textwidth]{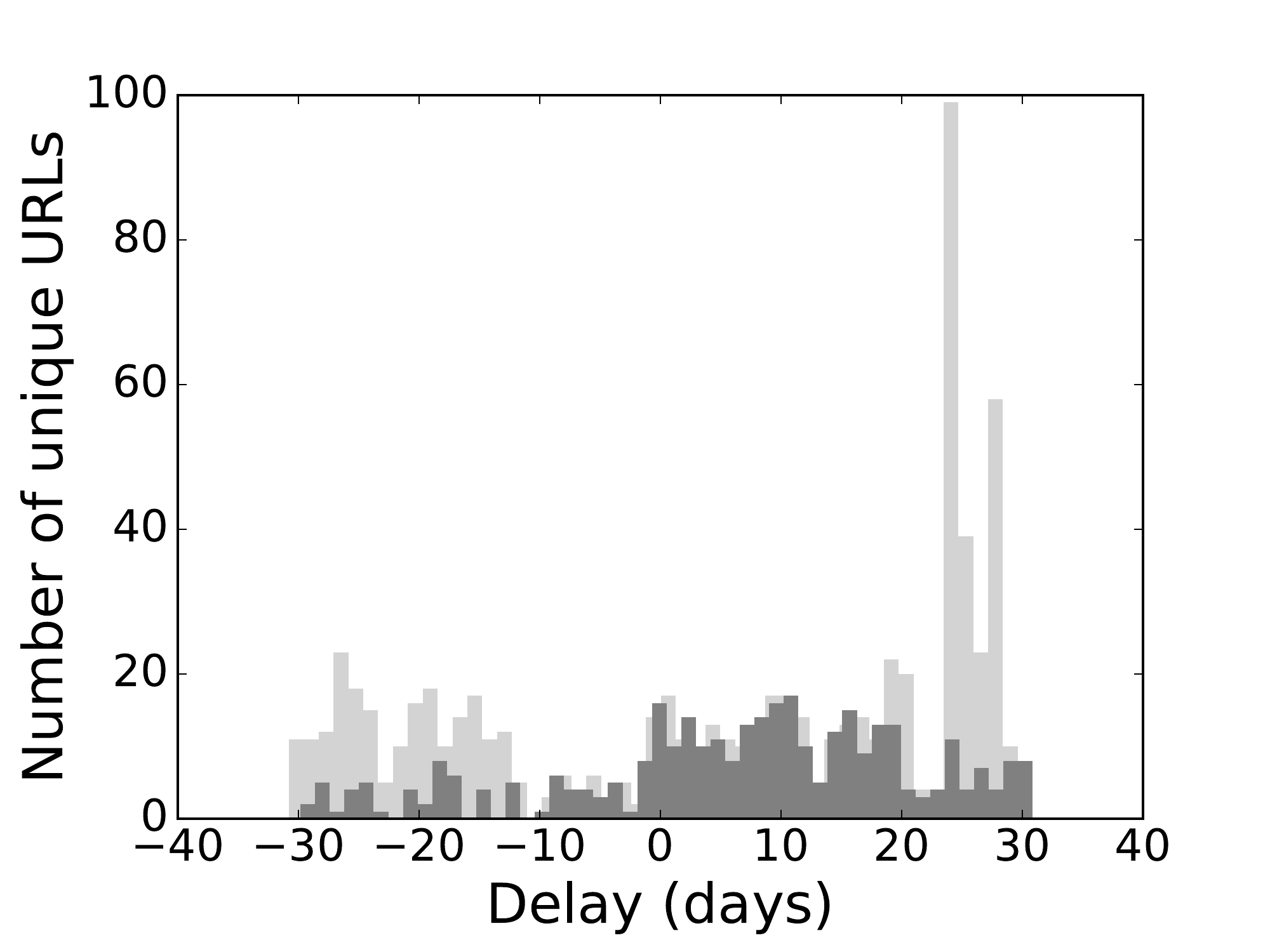}
        \label{fig:malware-urls-delay-histogram-oct-2017-top-domains-removed}
        }
    \subfloat[Malware Nov 2017]{
        \includegraphics[width=0.25\textwidth]{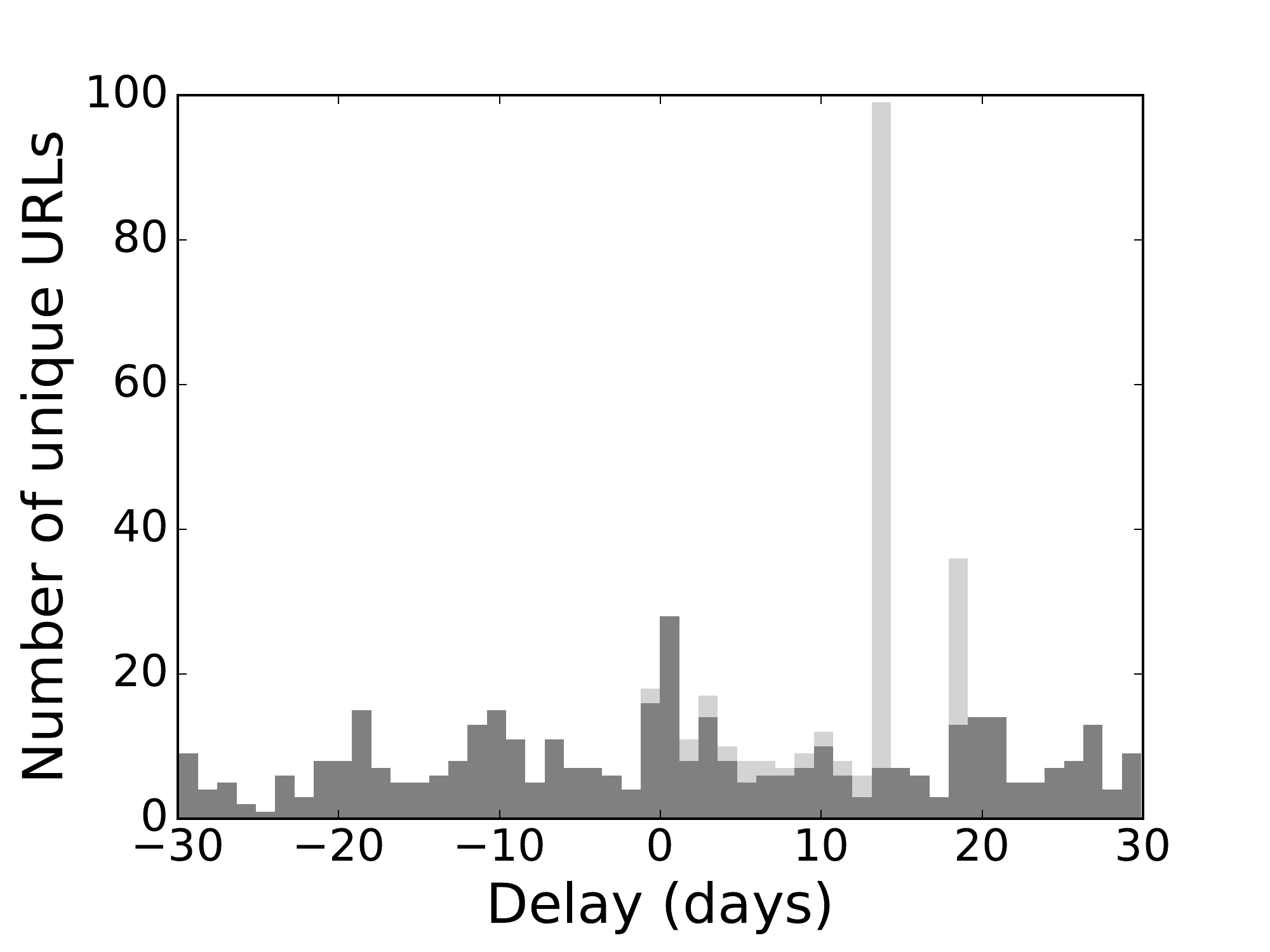}
        \label{fig:malware-urls-delay-histogram-nov-2017-minus-top-domains}
        }
        \caption{Delay from time of URL first tweet to appearing in GSB blacklist (including most frequent domain names) labelled social engineering and malware, November and October 2017.}
\end{figure*}

In this experiment we look at unique URLs that were first tweeted during October and November 2017 which were subsequently labelled as social engineering in the GSB blacklist within 1 month before or after being tweeted. Timestamps for when tweets were received from the Twitter Filter (URL) Stream are used as the tweet date and URL hash prefix timestamps from the GSB blacklist library are used to determine time first appeared in GSB blacklist to calculate total delay from tweet to blacklist, as described in the methodology section. During this experiment a total of 3,273 unique social engineering URLs were recorded in October and 930 in November. 

\begin{figure}[t!] 
\centering
    \subfloat[Social Engineering Oct 2017]{
        \includegraphics[width=0.25\textwidth]{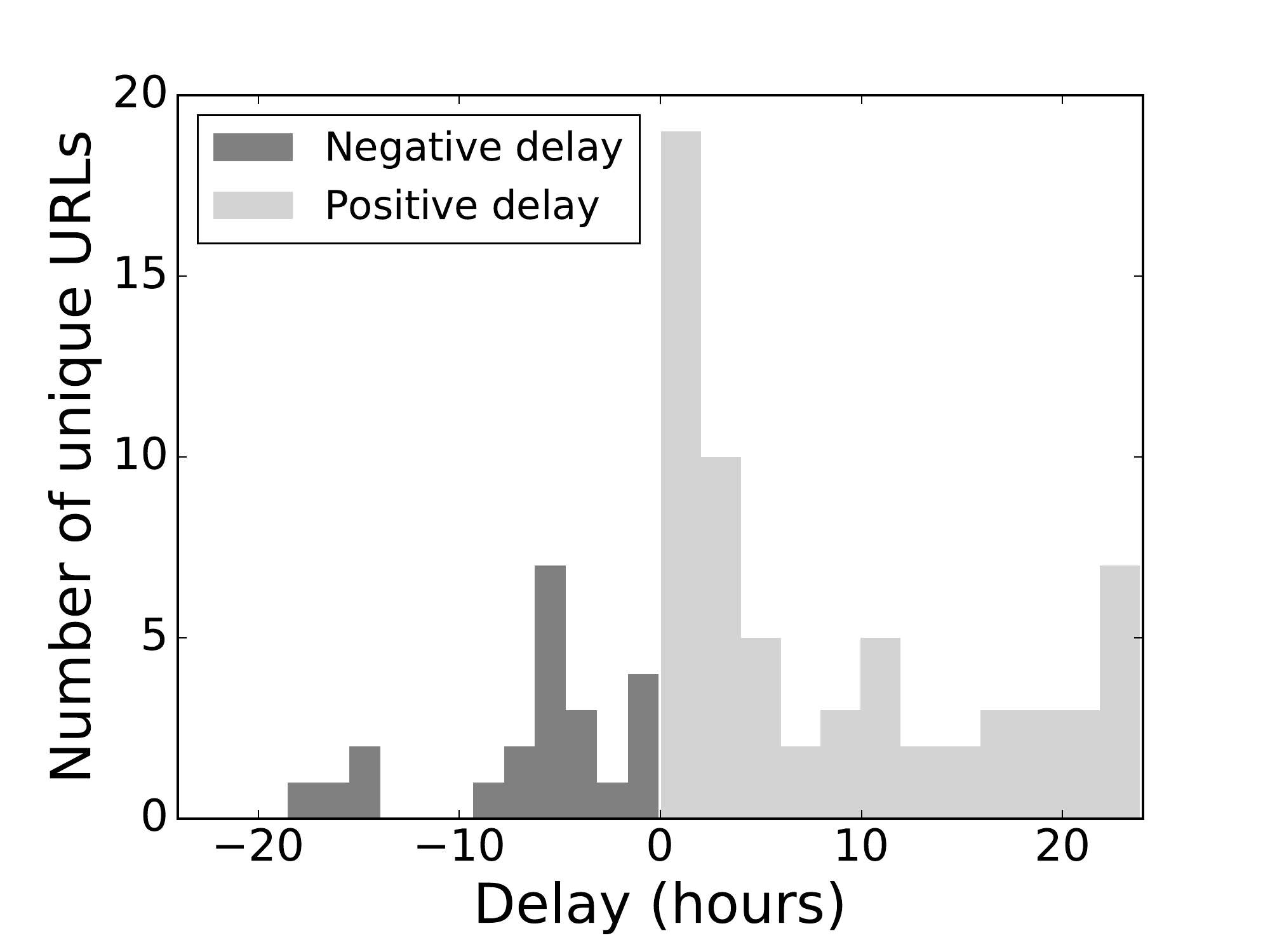}
        \label{fig:phishing-urls-delay-histogram-oct-2017-minus-top-7-domains-24-hours-12bins}
        }
    \subfloat[Social Engineering Nov 2017]{
        \includegraphics[width=0.25\textwidth]{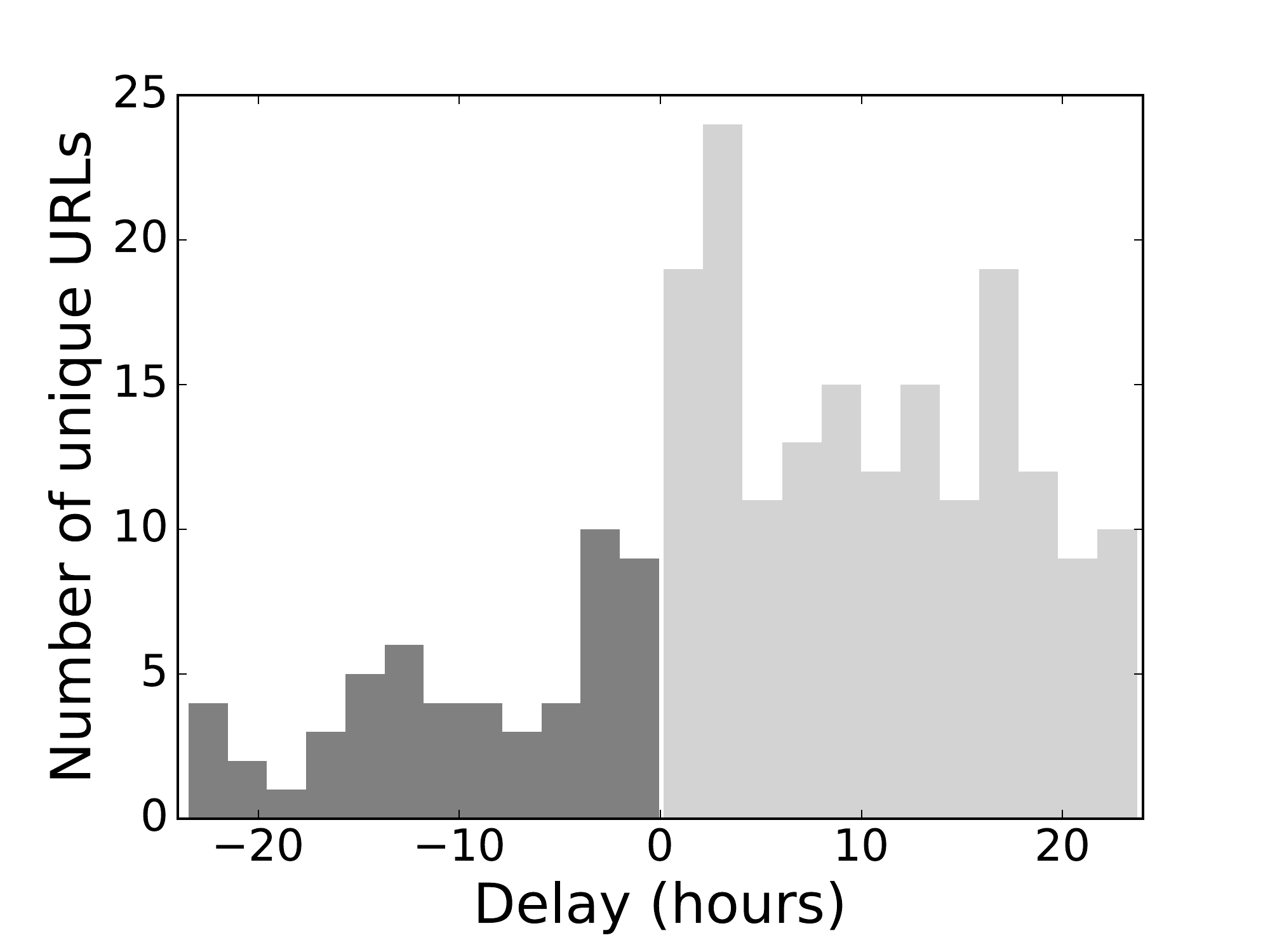}
        \label{fig:phishing-urls-delay-histogram-nov-2017-24-hours}
        }
        \caption{Social engineering URLs: delay from tweet to first appearing in GSB blacklist (Figures \ref{fig:phishing-urls-delay-histogram-oct-2017-minus-top-7-domains} and \ref{fig:phishing-urls-delay-histogram-nov-2017}), first 24 hours, October \& November 2017.}
\end{figure}

During October the majority of social engineering URLs saw a delay period of approximately 18 to 26 days from being tweeted to appearing in the GSB blacklist. Upon further investigation it was discovered that 7 domain names accounted for 76\% of the total dataset, 2,487 URLs, as shown in Table \ref{top-7-domains-oct}. All of the URLs contained within this dataset are \textit{HTTP}; none of them are \textit{HTTPS}. We extract the domain name for each URL per these examples: \textit{http://\textbf{example.com} /some-web-page.html}, \textit{http://subdomain. \textbf{example.com}}, \textit{https://\textbf{example.com} /some-secure-page.html} etc. Figure \ref{fig:phishing-urls-delay-histogram-oct-2017-minus-top-7-domains} shows frequency distribution for both the original 3,273 URLs along with the remaining 426 URLs after the top 7 domains names have been removed. This  histogram, with the top 7 URLs removed, shows that the majority of URLs appeared in the GSB blacklist within 6 hours of being tweeted, as seen, after being zoomed in to 24 hours, in Figure \ref{fig:phishing-urls-delay-histogram-oct-2017-minus-top-7-domains-24-hours-12bins}. A similar pattern is also seen in November where there is a peak at around 6 hours, as seen in Figure \ref{fig:phishing-urls-delay-histogram-nov-2017-24-hours}, although still a high number of URLs are blacklisted between 6 and 24 hours. Figures \ref{fig:phishing-urls-delay-histogram-oct-2017-minus-top-7-domains}-\ref{fig:malware-urls-delay-histogram-nov-2017-minus-top-domains} and \ref{fig:phishing-urls-delay-histogram-oct-2017-minus-top-7-domains-24-hours-12bins}-\ref{fig:phishing-urls-delay-histogram-nov-2017-24-hours} show the delay period between tweet and blacklist, with the number of unique URLs on the \textit{y} axis and delay period along the \textit{x} axis. A delay period greater than zero means that the URL appeared in the GSB blacklist after it appeared on Twitter. A delay of less than zero means that it was already in the GSB blacklist at time of Tweet. The negative delay values, in these graphs, show that large numbers of URLs were tweeted \textit{\textbf{after}} they appeared in the GSB blacklist. As with the previous subsection, this further suggests that Twitter are either not using the GSB blacklist or are allowing some URLs in GSB to be tweeted. This means that Twitter users are exposed to social engineering and malware attacks. 

In terms of the impact of these tweets, looking at just the top 7 most frequent domains that were first tweeted in October 2017 and appeared in GSB within 1 month before or after being tweeted, these 2,487 unique URLs were tweeted by 1,227 individual Twitter accounts, making up 4,930 total tweets. These 1,227 Twitter accounts have a combined number of 131,116,820 followers giving a sense of the total number of Twitter users potentially exposed to these social engineering tweets.

\begin{figure}[t!] 
    \subfloat[Phishing Oct 2017]{
        \includegraphics[width=0.25\textwidth]{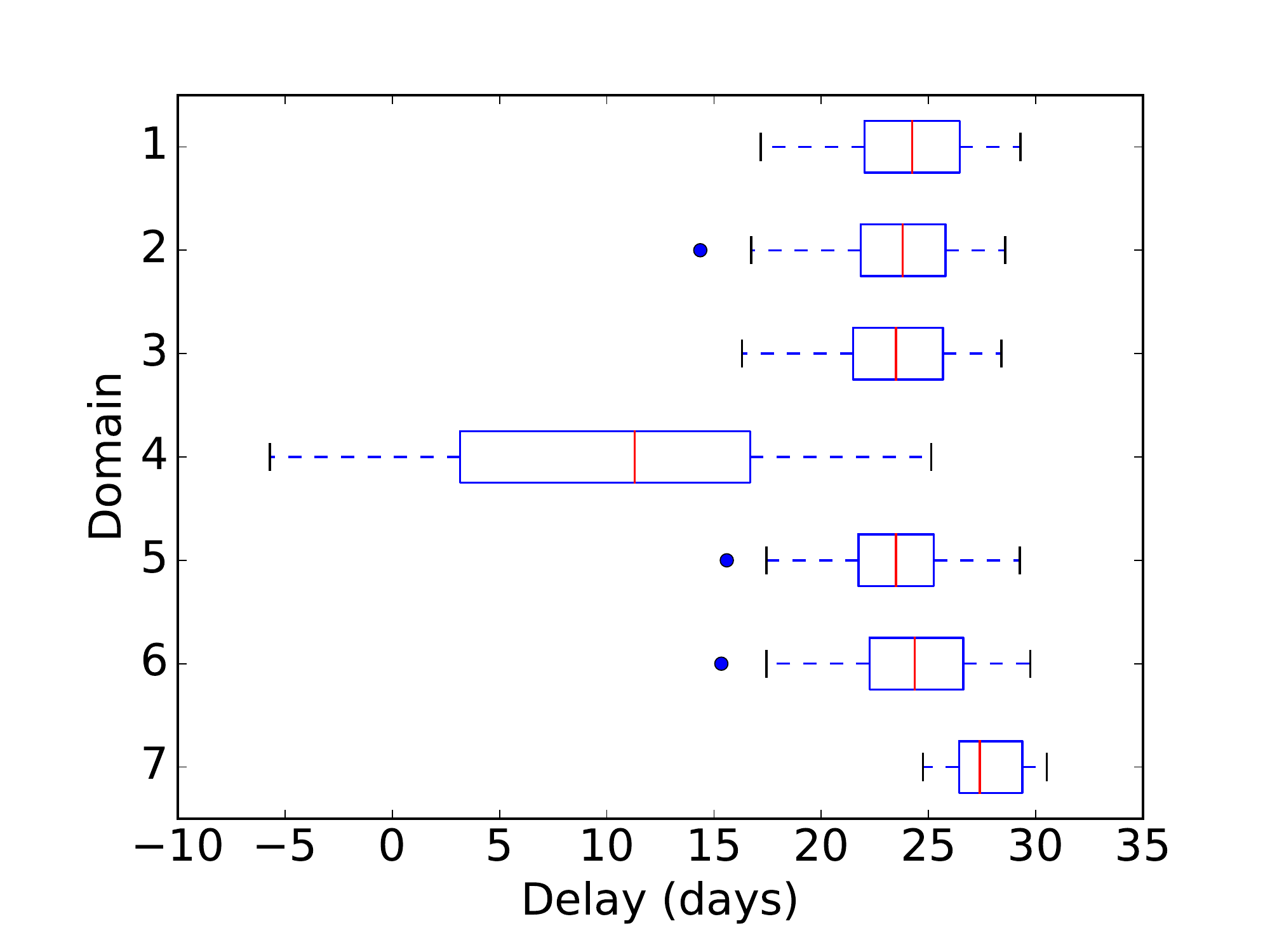}
        \label{fig:phishing-urls-delay-boxplot-oct-2017-top-7-domains}
        }
    \subfloat[Malware Oct 2017]{
        \includegraphics[width=0.25\textwidth]{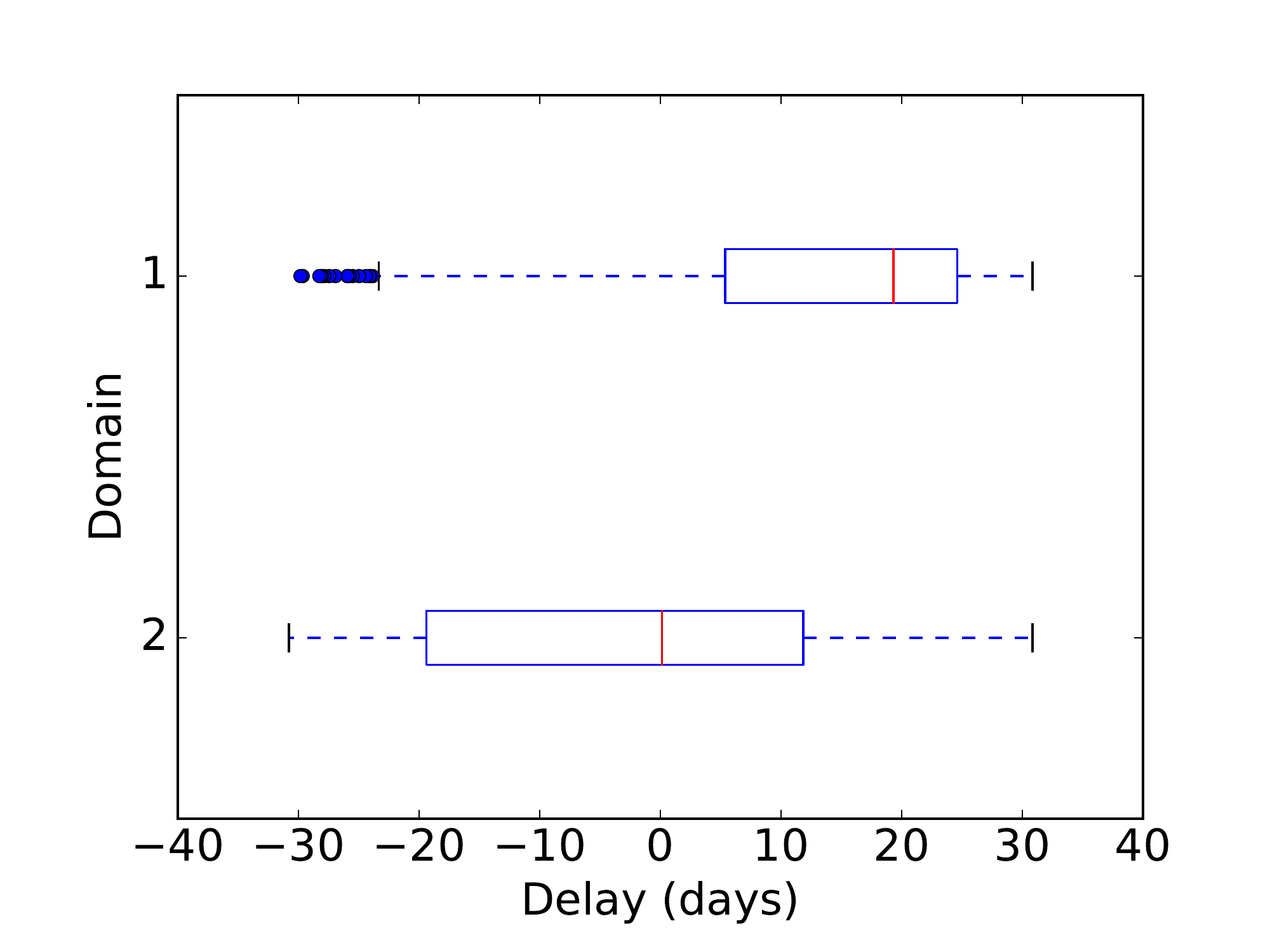}
        \label{fig:malware-urls-delay-boxplot-oct-2017-top-domains}
        }
        \caption{Box plots showing most frequent social engineering \& malware domains for October 2017.}
\end{figure}

\begin{table}[t!]
\begin{center}
\begin{tabular}{ |c|c|r| }
 \hline
   Domain & TLD & Number of URLs \\ 
 \hline
  1 & .cn & 614 \\  
 \hline
  2 & .cn & 582 \\ 
 \hline
  3 & .com & 554 \\
  \hline
  4 & .com & 273 \\ 
 \hline
  5 & .cn & 203 \\  
 \hline
  6 & .cn & 188 \\ 
 \hline
  7 & .life & 73 \\ 
 \hline
 \end{tabular}
\end{center}
\vspace{0.1cm}
\caption{October 2017 seven most frequent social engineering domains tweeted (domain names redacted).}
\label{top-7-domains-oct}
\end{table}

Figure \ref{fig:phishing-urls-delay-boxplot-oct-2017-top-7-domains} shows the distribution of these top 7 domains names in the dataset showing that Domains 1, 2, 3, 5 and 6 appear predominantly towards the 16 to 30 day delay mark with a few outliers around the 14 to 16 day mark, the majority of Domain 4 spans from -5 to 26 days and Domain 7 stays around the 25 to 31 day mark. 

When comparing the most frequently tweeted domains that are flagged as social engineering in GSB in October and November there are 4 domains names that appear in both months. This shows that, during the two months in which we collected data from Twitter and GSB, there were a number of large campaigns that spanned across both of these months. One of these domain names is in the Alexa top 100, suggesting that this website had become compromised, potentially by some sort of social engineering advert. One theory as to why Twitter continues to allow URLs from this domain, and others like it, to be tweeted on its network is because the main web browsers (such as Chrome, Safari, Firefox etc.) have built-in protection -- which should prevent users from visiting dangerous websites. Twitter can then outsource the protection of its users to the web browsers. This is also the case on both the Android and iOS Twitter apps whereby links are scanned by the Chrome and Safari web browser blacklists. One of the main weaknesses to this approach is that there may be an attack space when web browsers update their blacklists. If a user visits a newly blacklisted website, but their web browser has not updated their local copy of the blacklist, then the user will be allowed to visit the dangerous website without any warnings -- exposing them to the attack.

When analysing the target of the social engineering campaign tweets in October many of the tweets appear to be using click-bait techniques. These tweets often use misleading titles to promote, for example, health techniques with little evidence to backup their claims. Examples of tweets seen in our dataset include ``This Is What Happens When You Press This Point Near Your Ear For One Minute'' and ``This Leaking From Your Eye Can Be a Sign of a Dangerous Eye Infection''. These click-bait techniques are commonly used to attract large numbers of people to a website in order to generate revenue from adverts. 

We then repeat the experiment, only this time analysing all tweeted malware URLs, as classified by GSB. A total of 718 unique malware URLs were recorded during the same timeframe in October 2017 and 543 unique malware URLs in November 2017. When looking at the frequency distribution of delays, the largest peak of GSB blacklisted malware URLs in October occurs at approximately 25 days and was caused by 2 domain names (consisting of 219 and 161 URLs) making up 39\% of the October dataset of 718 URLs. The total number of tweeted Malware URLs in October can be seen in Figure \ref{fig:malware-urls-delay-histogram-oct-2017-top-domains-removed} and shows frequency distribution for the month including the 2 outlying domain names. Figure \ref{fig:malware-urls-delay-boxplot-oct-2017-top-domains} shows the distribution of these top 2 domain names in the dataset, showing that Domain 1 mostly covered days 5 to 25, with its median at approximately 19.5 days. Domain 2 is predominantly spread over the -20 to 12 day delay period, with its median being just over 0 days. Finally, in November 2017, the largest peak appears at around 14 days and is caused by 1 outlying domain name (consisting of 140 URLs) which made up 26\% of the dataset. The frequency distribution for November can be seen in Figure \ref{fig:malware-urls-delay-histogram-nov-2017-minus-top-domains}.

\begin{figure}[t!]
\centering
\includegraphics[width=0.5\textwidth]{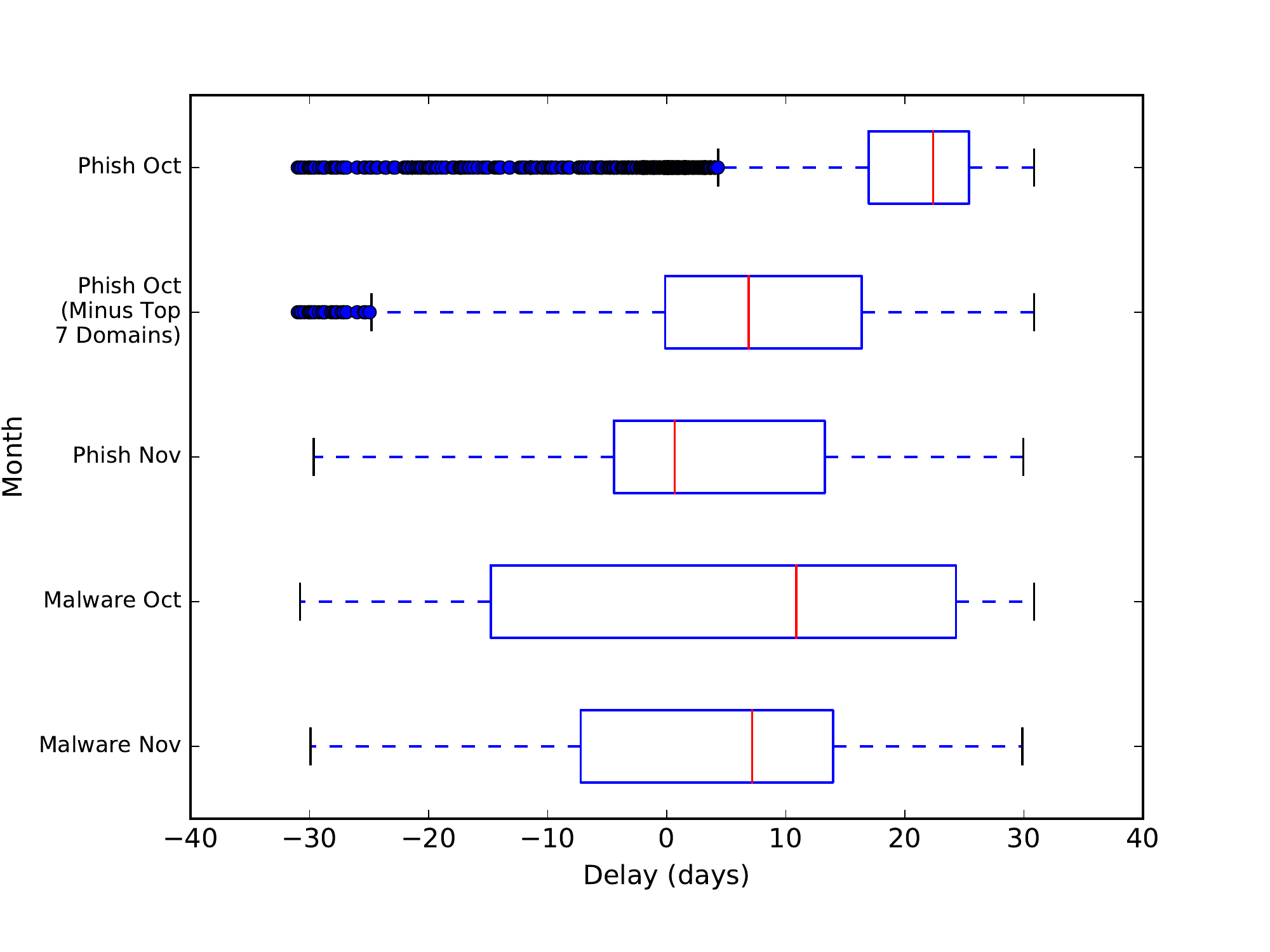}
\caption{Delay from first tweet to first appearing in GSB blacklist -- social engineering and malware, October and November 2017.}
\label{fig:phishing-malware-urls-delay-boxplot-oct-nov-2017}
\end{figure}

\begin{figure*}[htp] 
    \subfloat[Social Engineering Oct 2017]{
        \includegraphics[width=0.25\textwidth]{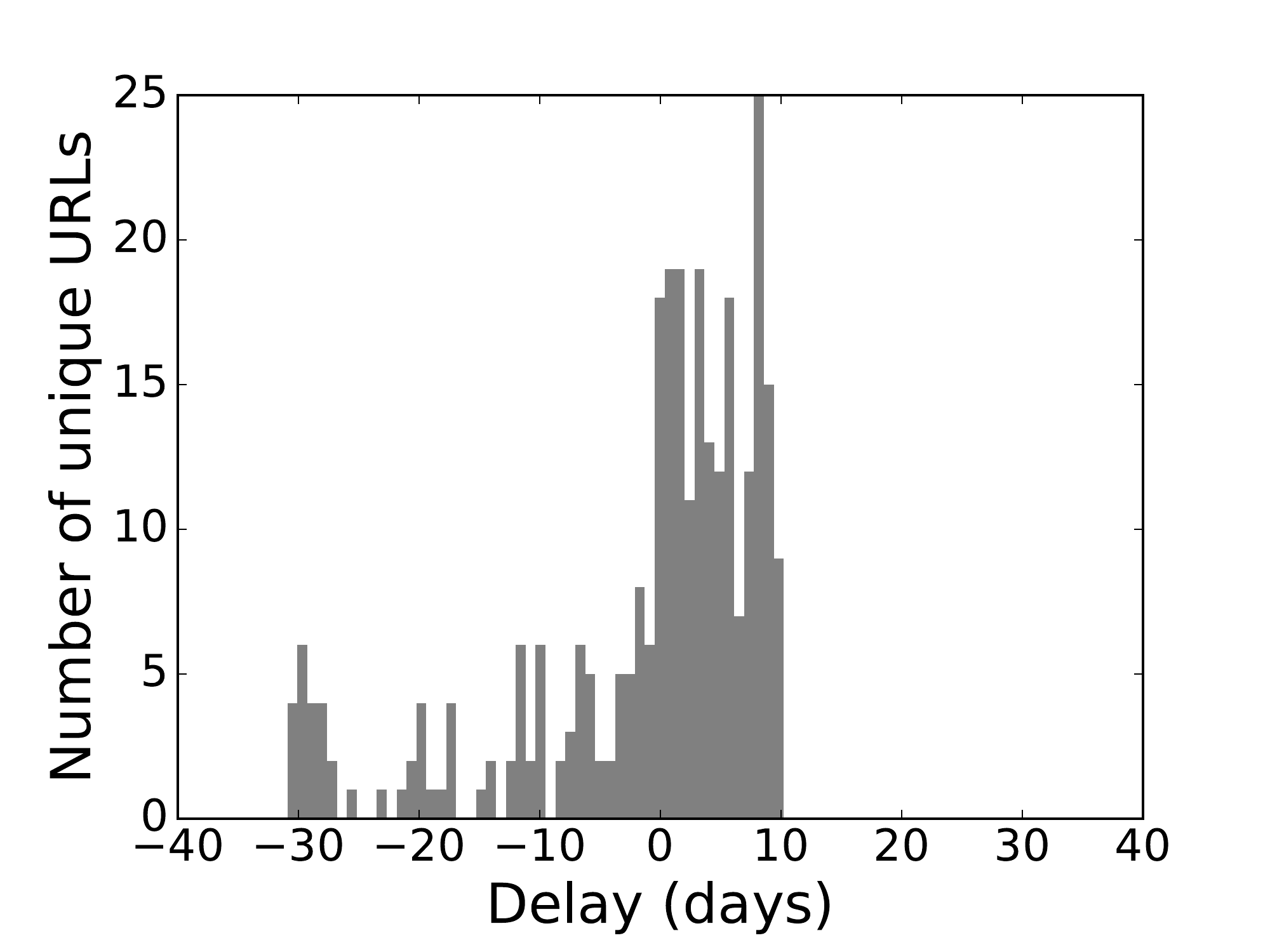}
        \label{fig:phishing-urls-delay-histogram-oct-2017-twitter-search-api}
        }
    \subfloat[Social Engineering Nov 2017]{
        \includegraphics[width=0.25\textwidth]{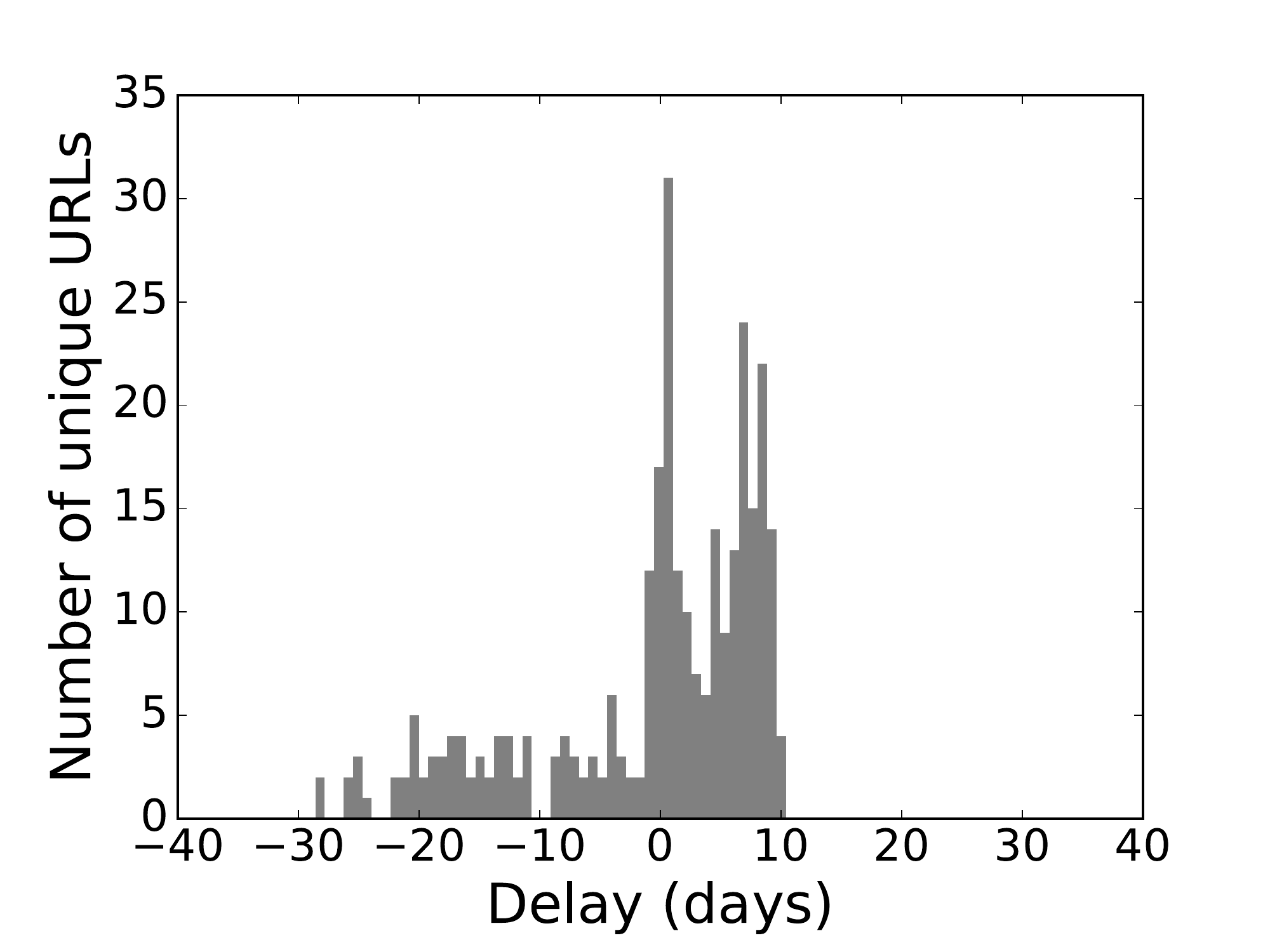}
        \label{fig:phishing-urls-delay-histogram-nov-2017-twitter-search-api}
        }
    \subfloat[Malware Oct 2017]{
        \includegraphics[width=0.25\textwidth]{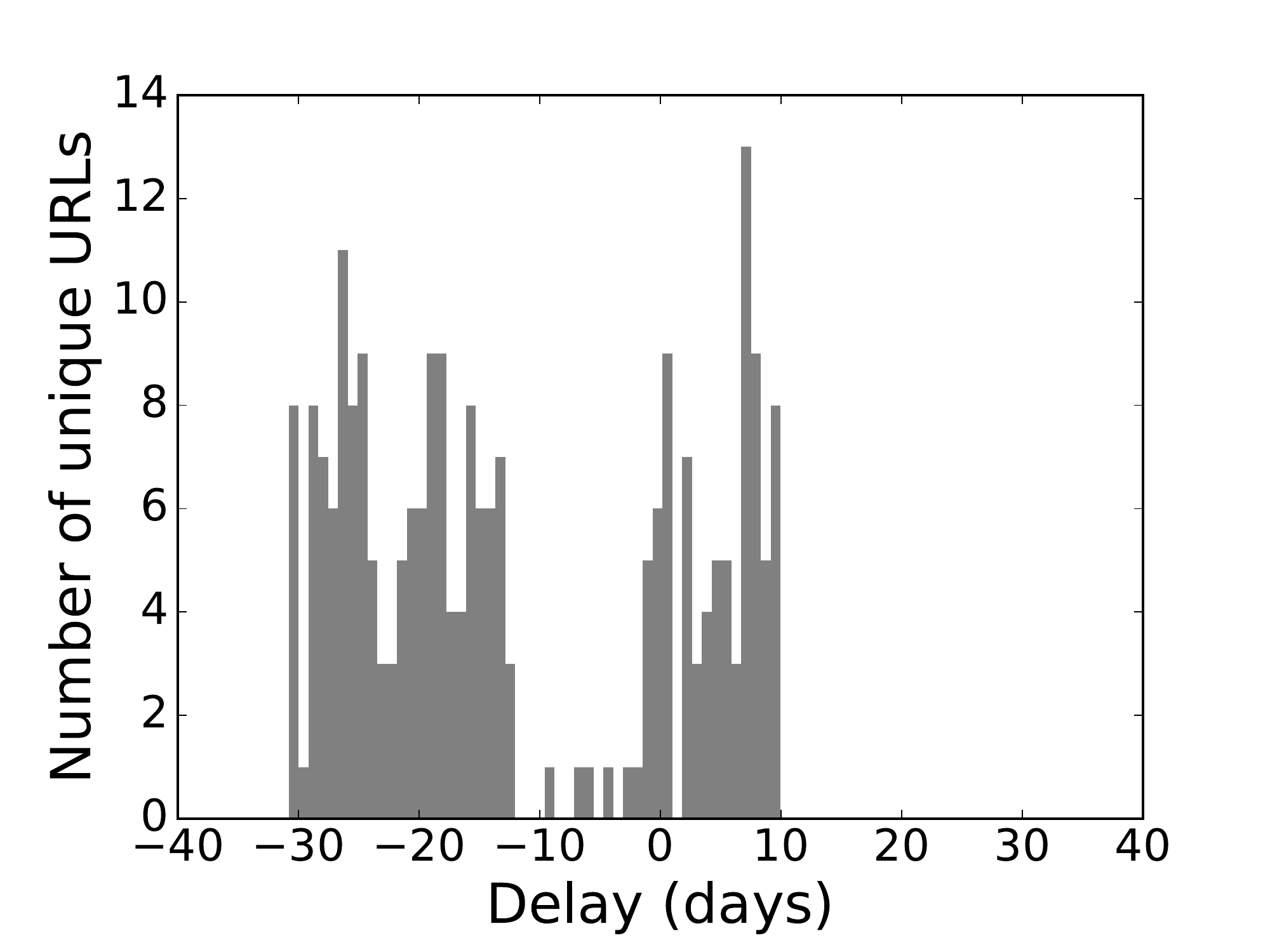}
        \label{fig:malware-urls-delay-histogram-oct-2017-twitter-search-api}
        }
    \subfloat[Malware Nov 2017]{
        \includegraphics[width=0.25\textwidth]{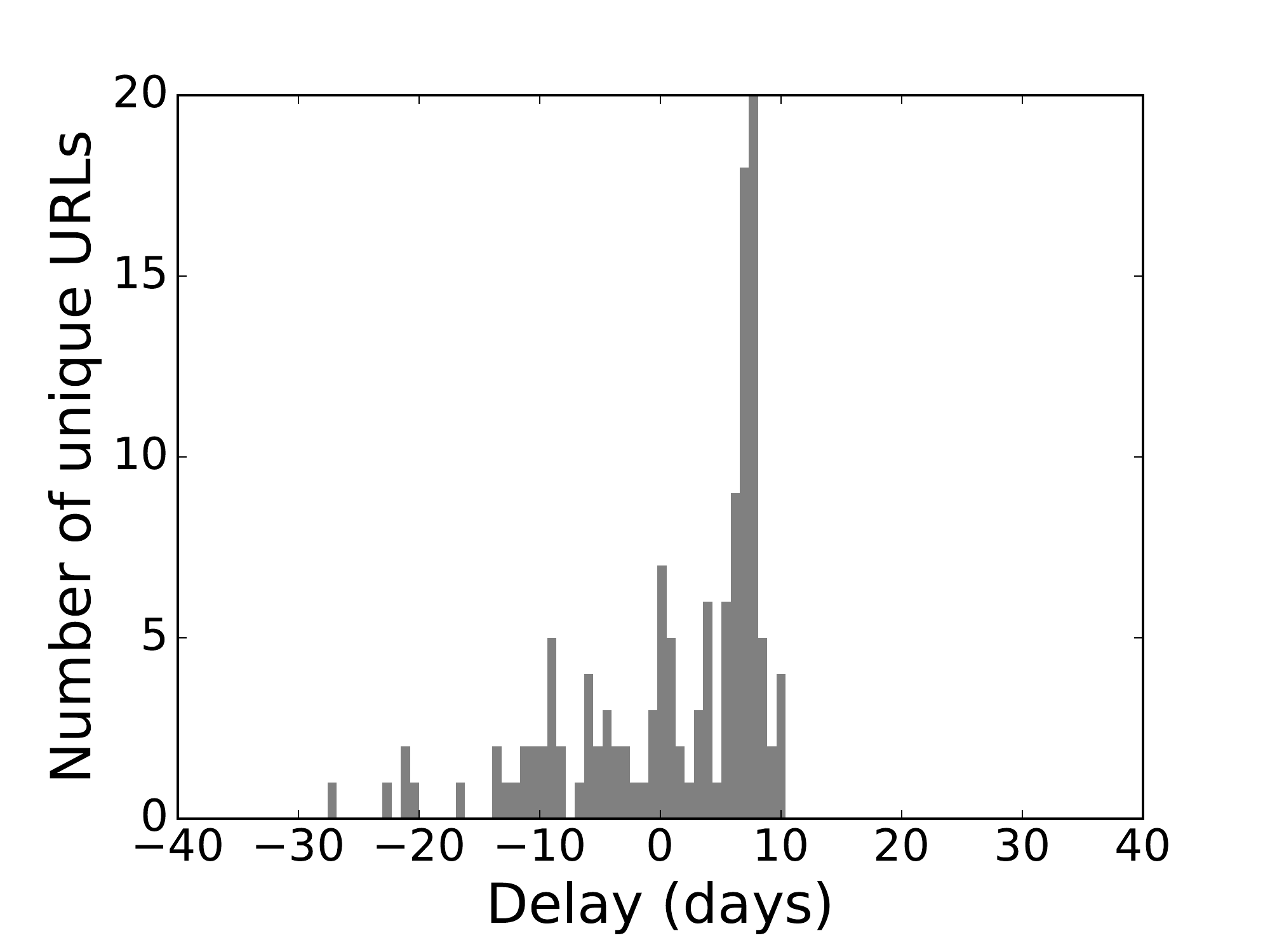}
        \label{fig:malware-urls-delay-histogram-nov-2017-twitter-search-api}
        }
        \caption{Delay from first tweet to first appearing in GSB blacklist -- using Twitter Search API to determine URL first tweet date -- social engineering and malware, October and November 2017.}
\end{figure*}

Figure \ref{fig:phishing-malware-urls-delay-boxplot-oct-nov-2017} shows a boxplot of tweeted GSB blacklisted social engineering and malware URL delays in October and November 2017. The first row shows the distribution of social engineering URLs in October 2017, before the top 7 domains were removed whilst the second row shows the same timeframe but with the top 7 domains removed. Row three shows GSB blacklisted social engineering URLs in November, row four shows malware URLs in October and row five shows malware URLs in November 2017. This shows that, for social engineering URLs, the median delay time was around 7 days in October and just over 0 days in November. For malware URLs the median delay was around 11 days in October and around 8.5 days in November. 

These graphs show that GSB appears to be quicker at detecting social engineering websites than malware websites. One reason for this may be that the criteria for the social engineering flag may include a wider net. Therefore, as we saw with the Alexa top 100 domain, some high traffic websites may become blacklisted when they fall into this net. Whereas flagging a website in the malware blacklist requires Google to be certain the website is harming - or attempting to harm - the user's computer in some way. This potentially stricter classification may take more time to confirm and may explain why malware is slower to detect than social engineering in our results.

\paragraph{\textbf{Findings:}} One of the key takeaways from these experiments is that Twitter allow considerably more URLs to be tweeted \textbf{\textit{after}} appearing in the GSB blacklist, compared to the 2010 study \cite{grier2010spam}. As previously mentioned, this may be because Twitter is relying more on web browsers' built-in protection from malware and phishing URLs. However, one of the biggest weaknesses to this approach is that the built-in blacklists used by web browsers take time to update and this creates an attack space. The results in this section also show there is a significant delay -- 20 to 30 days in some cases -- before URLs are blacklisted. We also see where a combined total of 131,116,820 Twitter users are exposed to 2,487 unique blacklisted URLs. This means Twitter users are exposed to these dangerous attacks for a substantial amount of time.

Even though the experiments in this subsection do not identify absolute earliest time of tweet, our delay measurement will always be an underestimate. Therefore the real situation, in terms of Twitter users being exposed to dangerous URLs due to blacklist delay times, is much worse. This methodological weakness is addressed in the next subsection.

\subsection{Blacklist Delays -- Twitter Search API}
\label{sec:results-twitter-search}

The experiments in this subsection aim to further improve the accuracy of the experiments carried out in the previous subsection. We do this by making use of Twitter's Search API to determine the original tweet date for blacklisted URLs. Also, because we use Twitter's 1\% feed of tweets there may be instances where a URL appears outside of our Twitter Stream. By using Twitter's Search API we can determine when a given URL was tweeted. The measurements taken in this experiment are the same as in the previous subsection, that is the delay between a blacklisted URL first being tweeted and first appearing in the GSB blacklist within 1 month before of after tweet date, only in this section each URL is searched for on Twitter and the timestamp of that search result is used for the delay calculation. Using this method, in October 2017, 295 social engineering and 230 malware URLs are recorded; their delays can be seen in Figures \ref{fig:phishing-urls-delay-histogram-oct-2017-twitter-search-api} and \ref{fig:malware-urls-delay-histogram-oct-2017-twitter-search-api}. In November 2017, 284 social engineering and 131 malware URLs are recorded and can be seen in Figures \ref{fig:phishing-urls-delay-histogram-nov-2017-twitter-search-api} and \ref{fig:malware-urls-delay-histogram-nov-2017-twitter-search-api}. 

It is important to note that there are significantly fewer URLs in this part of the dataset. This is because Twitter states that its Search API is not a complete search, therefore some URLs we try to determine original tweet timestamps for cannot be found. In this case these URLs are dropped from the dataset. A clear pattern that emerges in all 4 of these graphs is that there are no URLs with a delay from first tweet to first blacklist of more than 10 days. This is because Twitter's Search API is limited to 7-10 days; any URLs the system searches for that appeared in the GSB blacklist more than 7-10 days after being tweeted will not show up in a Twitter search if the URL has not been tweeted again since. This limits these graphs, since they show a reduced picture of delays between URLs being tweeted and appearing in the GSB blacklist. However, as seen in the previous two sections, there are still high numbers of URLs already in the GSB blacklist at time of tweet. 

\paragraph{\textbf{Findings:}} Measurements in the previous subsection do not show the worst-case scenario in terms of delay from first tweet to appearing in blacklist because URLs may have been previously tweeted. However, results in this section, whilst showing fewer URLs in the dataset, do show the worst case scenario for delay from first tweet to blacklist membership. This adds additional evidence that Twitter are not blocking all GSB URLs and may to be relying on other, possibly third-party techniques, to protect its users against attacks. There is also a significant number of URLs that take between 0 and 10 days to appear in the GSB blacklist -- meaning users are exposed to social engineering and malware attacks during these delay periods.

\subsection{Blacklisted URL Clicks}

\begin{table}[t!]
\small
\begin{center}
\begin{tabular}{ | p{2cm} | r | r | r | r | } 
 \hline
   & \multicolumn{2}{|c|}{GSB SE} & \multicolumn{2}{|c|}{GSB Malware} \\ 
  \hline
   & Oct & Nov & Oct & Nov \\ 
  \hline
  Total tweets containing Bitly URLs & 1126 & 146 & 32 & 103 \\  
  \hline
   Total unique Bitly URLs & 376 & 141 & 30 & 66 \\  
  \hline
   Percentage of all blacklisted URLs in this category and timeframe & 11\% & 15\% & 4\% & 12\% \\  
  \hline
  Total Bitly clicks & 991,012 & 450,039 & 61,140 & 194,503 \\  
  \hline
\end{tabular}
\end{center}
\vspace{0.1cm}
\caption{Total number of tweets containing Bitly URLs, total number of unique Bitly URLs, percentage of all URLs for each category and timeframe, and total Bitly clicks for tweets containing GSB blacklisted social engineering (SE) and malware URLs, in our dataset, in October and November 2017.}
\label{tab:bitly-click-stats}
\end{table}

To explore the impact of tweets that contain blacklisted URLs, we lookup Bitly URLs that either directly appear in or are embedded in the redirection chain that leads to the GSB blacklist, in our dataset. Bitly~\cite{bitly} is a URL shortening service that also provide public analytics for URL clicks, referrers, and location, via an API. By extracting Bitly links from our dataset of tweeted URLs that subsequently appear in the GSB blacklist, we can then use the Bitly API to lookup how many clicks each URL received.

Table \ref{tab:bitly-click-stats} shows, from our dataset of tweeted URLs that subsequently appeared in the GSB blacklist, the total number of unique Bitly URLs, percentage of all URLs in this category and timeframe, total number of tweets containing Bitly URLs for this category and timeframe, and total number of Bitly URL clicks, during October and November 2017. In October, there were 376 unique Bitly URLs that were either flagged themselves or part of a redirection chain that was in the GSB blacklist as social engineering. These 376 Bitly URLs make up 11\% of the 3,273 total social engineering URLs detected in that month in our dataset. The total number of clicks for this 11\% is 991,012. 

To investigate the impact of tweeting a blacklisted URL to a Twitter account with a high number of followers, we extracted a blacklisted URL, from our dataset, that uses Bitly. The blacklisted Bitly URL was tweeted by an account with 3.7 million followers on October 24 and flagged as social engineering in GSB on November 11. The URL received 276 clicks during the week of October 22 2017, of which 270 came from Twitter. 176 of these clicks came from the USA, 19 from Canada, 12 from the UK and the remaining 34 from elsewhere. This URL did not receive any more clicks after the week of October 22 at which point it appears to have been blocked by Bitly. This example shows that a single tweet, from a high follower account, posting a dangerous URL, can receive a high number of global clicks -- therefore exposing a large amount of Twitter users to the attack. It also shows that GSB took 18 days to add the URL to its blacklist, while Bitly appears to have blocked the URL much sooner. In this scenario, Twitter appears to have outsourced its filter to Bitly -- relying on Bitly to protect Twitter's own users.

\paragraph{\textbf{Findings:}} These results show that, in one month alone, 1,052,152 clicks were exposed to dangerous malware and social engineering attacks due to Twitter not blocking these harmful URLs. These click metrics represent 11\% of our dataset, which is, approximately, 1\% of all global tweets on Twitter -- giving a sense of the scale and impact caused by Twitter allowing blacklisted URLs to appear on their social network.

\subsection{Posting Blacklisted URLs to Twitter}
In a separate experiment we created a private account on Twitter whereby the account's tweets were not publicly visible. We then attempted to tweet a sample of 30 blacklisted URLs: 10 from GSB, 10 from Open Phish and 10 from Phish Tank. In this experiment, 8 of the Open Phish URLs and 9 of the Phish Tank URLs could not be posted to Twitter. All of the GSB URLs were posted successfully to Twitter. For tweets containing blacklisted URLs that could not be posted to Twitter this error message was displayed: ``This request looks like it might be automated. To protect our users from spam and other malicious activity, we can't complete this action right now. Please try again later''. We were able to tweet messages that did not contain blacklisted URLs without receiving this error message. This suggests that Twitter may display this generic error message when URLs that it has filtered are requested to be tweeted on the social network. It is important to note that this was a small-scale study and that the Twitter account used for this experiment was set to private, therefore all tweets were hidden from the public. Public Twitter accounts may see different results in this experiment -- for example: public tweets may go through a stricter filtering process. Due to ethical considerations, we did not post any public tweets containing blacklisted URLs. 

\paragraph{\textbf{Findings:}} The outcome of this experiment shows that Twitter appears to be blocking more URLs on the Phish Tank and Open Phish blacklists compared to GSB. Providing further evidence that Twitter is not using the GSB blacklist  -- therefore exposing users to dangerous URLs.

\subsection{URL Time in GSB}

\begin{figure*}[htp] 
    \subfloat[Social Engineering Oct]{
        \includegraphics[width=0.25\textwidth]{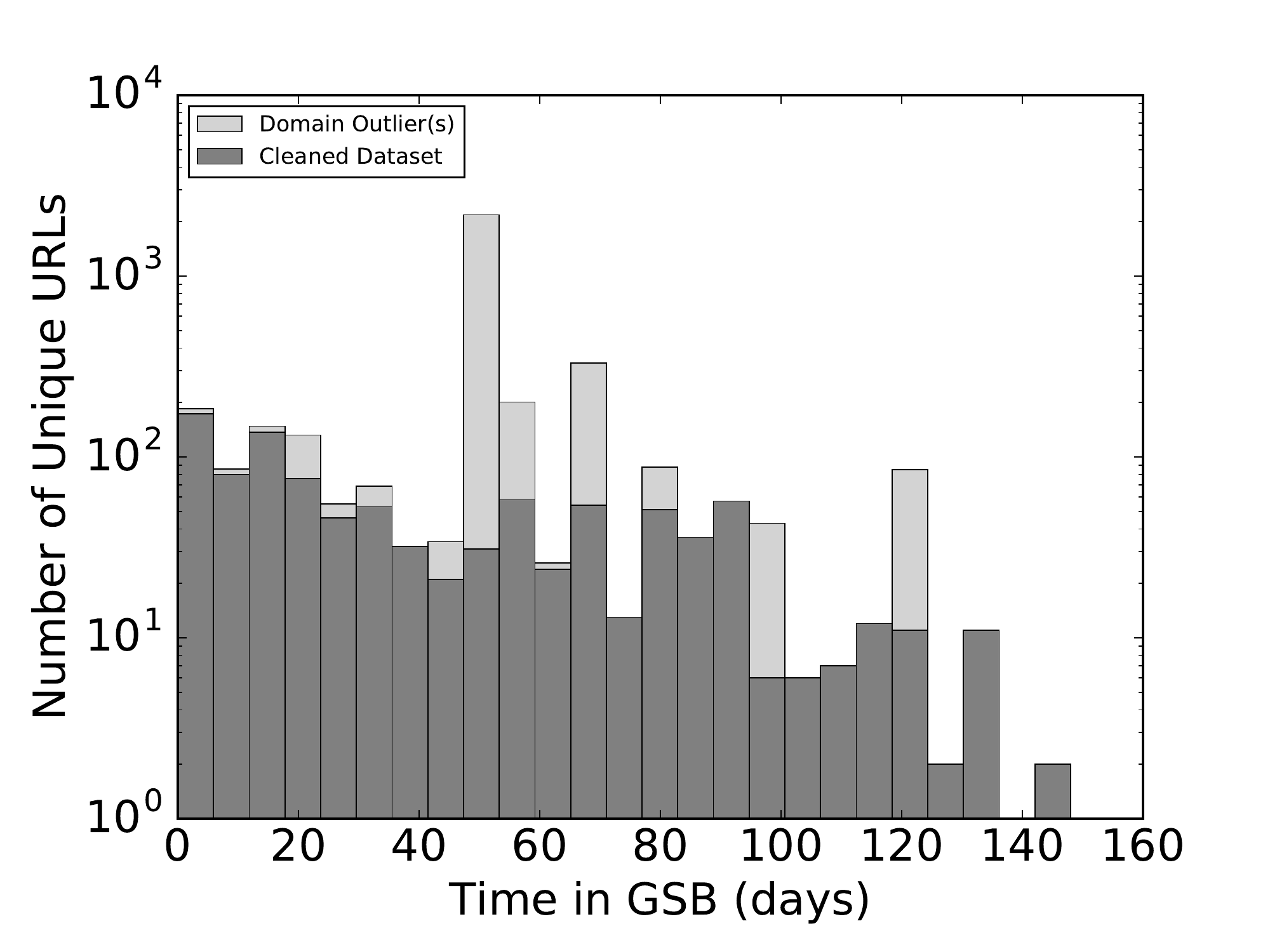}
        \label{fig:phishing-urls-dropout-stats-oct-2017}
        }
    \subfloat[Social Engineering Nov]{
        \includegraphics[width=0.25\textwidth]{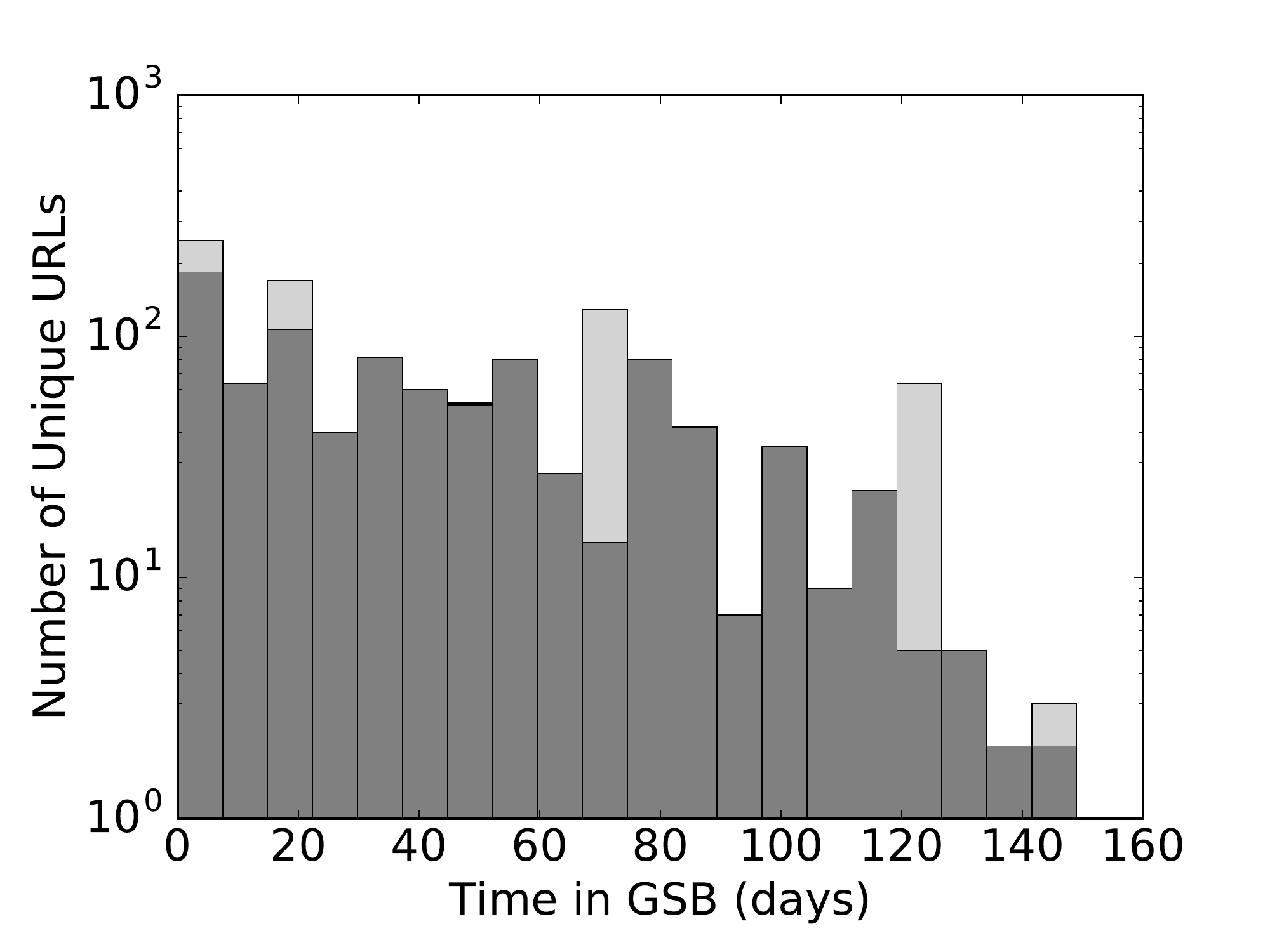}
        \label{fig:phishing-urls-dropout-stats-nov-2017}
        }
    \subfloat[Malware Oct]{
        \includegraphics[width=0.25\textwidth]{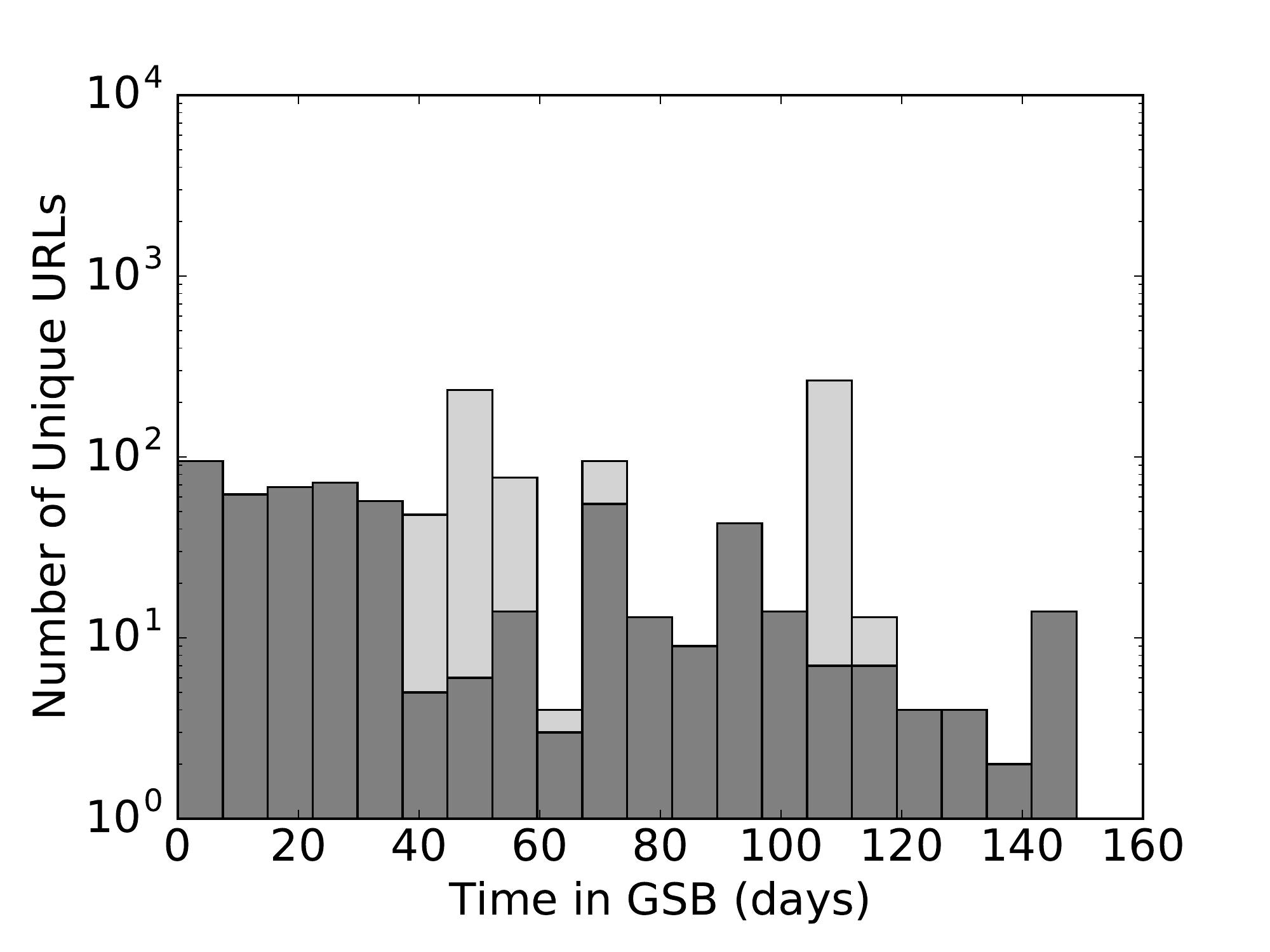}
        \label{fig:malware-urls-dropout-stats-oct-2017}
        }
    \subfloat[Malware Nov]{
        \includegraphics[width=0.25\textwidth]{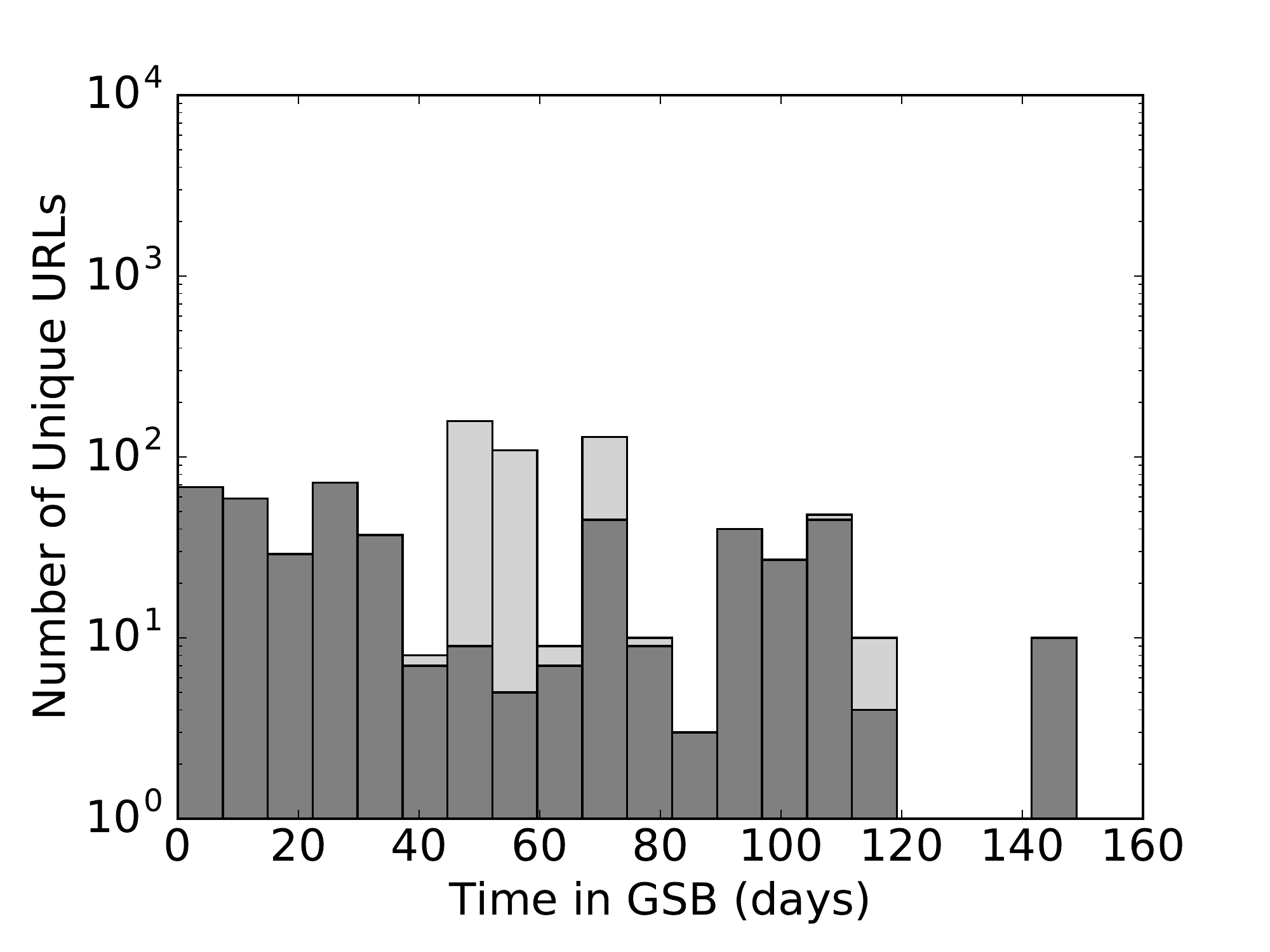}
        \label{fig:malware-urls-dropout-stats-nov-2017}
        }
        \caption{Unique social engineering \& malware URLs duration in GSB -- first tweeted October, November 2017.}
\end{figure*}

This section explores the duration of time that unique URLs remained in the GSB blacklist for. Each experiment takes all unique URLs that were first tweeted in a given month, then, if a URL is not in GSB at time of tweet, the duration in GSB is calculated as when the system first detects the URL in GSB to when the system last saw the same URL in GSB. If a URL is already in GSB at time of tweet then the GSB library URL hash prefix timestamp is used as time first blacklisted and the time our system last saw the URL in GSB as the final timestamp. The difference between these timestamps is used to calculate total time in GSB for each URL. These duration periods are then plotted on histograms to show the frequency of different duration in GSB for all URLs.

Figures \ref{fig:phishing-urls-dropout-stats-oct-2017} and \ref{fig:phishing-urls-dropout-stats-nov-2017} show the duration of time that social engineering URLs spent in the GSB blacklist in October and November 2017 and Figures \ref{fig:malware-urls-dropout-stats-oct-2017} and \ref{fig:malware-urls-dropout-stats-nov-2017} show the duration of time malware URLs appeared in the GSB blacklist in October and November 2017. All four of these graphs have a logarithmic scale on the \textit{y} axis so both high and low numbers are illustrated clearly. 

\paragraph{\textbf{Findings:}} One of the main conclusions from these graphs is that there is a general downward trend. This shows that, over time, the number of URLs in the GSB blacklist is reducing. This means that URLs are removed from the blacklist, presumably once they are no-longer a threat. Our experiment ran for 150 days and there were over 1,000 URLs remaining in the blacklist, for each category, at the end of the experiment -- meaning that many URLs remained in the GSB blacklist for at least 150 days. Some of these URLs may still be dangerous, however, there may be false positives in this blacklist which would mean these URLs are, unnecessarily, being blocked. Exploring long-term false positives in GSB is something we may explore in future work.

\section{Discussion}
\label{sec:discussion}

\subsection{Limitations}
\label{sec:limitations}

Twitter's Search API is limited to 7-10 days and is not a complete search, therefore the resulting dataset in Section \ref{sec:results-twitter-search} is reduced. Despite this, the methodology increases  accuracy of both the dataset and results in Section \ref{sec:results-time-of-first-tweet}. Thereby producing the worst case scenario result, from the perspective of users, when calculating delay from first tweet to first appearing in blacklist. 

Twitter's approximately 1\% data stream provides a reduced dataset, therefore limiting the determination of original URL tweet timestamps (i.e. if a URL is tweeted outside the data stream). We compensate as much as possible for this by using techniques such as Twitter's Search API to determine first tweet timestamps.

Our study may capture benign websites that became compromised. In future work we may explore compromised websites further, for example, by analysing percentage of compromised versus attack websites in our results (using GSB's terminology).

GSB uses path prefix expansion; iteratively trying broader and broader URLs (e.g., x.y.z/a/b/c, x.y.z/a/, x.y.z, y.z). This could result in newly blacklisted hosts, from fresh incidents, being misinterpreted as missed historical URLs. Potential mitigation could involve GSB library modification to flag if entire domains become blacklisted. Results could then exclude blacklisted domains.

We do not detect tweets containing phishing or malicious URLs that never make it into GSB. Therefore GSB is our ``ground truth''. We attempt to mitigate this by using the Open Phish and Phish Tank blacklists.

\subsection{Twitter Filter Analysis}
\label{sec:twitter-filter-analysis}

We have hypothesised that Twitter may have developed their own method to filter dangerous URLs from their network, to protect their users, and are no longer using the GSB blacklist. Whilst we could carry out experiments to analyse this further, we would essentially be ``reverse engineering'' Twitter's filtering process. It is hard to do this without violating Twitter's terms of use.

\section{Conclusion}
\label{sec:conclusion}

This paper examined how effective URL blacklists are in protecting Twitter users against phishing and malware attacks. We analysed over 182 million URL-containing public tweets collected from Twitter's Stream API, over a 2 month period, and compared these URLs against 3 popular social engineering, phishing, and malware blacklists. Our main discovery was that, although the majority of phishing and malware URLs are detected by the GSB blacklist (which is used by popular web browsers) within 6 hours of being tweeted, there are still a large number of URLs that take at least 20 days to appear in GSB. We discovered 4,930 tweets containing URLs leading to social engineering websites that took between 18 and 30 days to appear in the blacklist. Between them, these 4,930 tweets had been tweeted to over 131 million Twitter users. We also discovered 1,126 tweets containing 376 blacklisted Bitly URLs that had received a combined total of 991,012 clicks. These URLs represented 11\% of the total blacklisted social engineering URLs in that month. The fact that the GSB blacklist can take weeks to detect dangerous URLs poses serious security risks to Twitter users: tweets containing blacklisted URLs are sent to large numbers of followers and receive a significant amount of clicks, thereby exposing users to dangerous websites. Conversely, and surprisingly to us, there are large numbers of URLs being tweeted that have already been blacklisted by GSB. This strongly indicates that Twitter is not using the GSB blacklist to block malicious tweets at the time of tweeting, contrary to what was once reported to be the case~\cite{zdnet}. In summary, whilst blacklists are reasonably effective at protecting Twitter users from phishing and malware attacks, there is still an unprotected space that leaves Twitter users vulnerable.

\bibliographystyle{ACM-Reference-Format}
\bibliography{catch_me_on_time_arXiv}

\end{document}